\documentclass[11pt]{article}

\def\viewchanges{1}

\usepackage[utf8]{inputenc} 
\usepackage[T1]{fontenc}    
\usepackage{url}            
\usepackage{booktabs}       
\usepackage{amsfonts}       
\usepackage{nicefrac}       
\usepackage{microtype}      
\usepackage{lipsum}
\usepackage{graphicx}
\usepackage{amsmath}
\usepackage{amsthm}
\usepackage{caption}
\usepackage{subcaption}
\usepackage{enumerate}
\usepackage{algorithmic}
\usepackage{algorithm}
\usepackage{lipsum}
\usepackage{multirow}
\usepackage{array}
\usepackage{multicol}
\usepackage{wrapfig}
\usepackage{float}
\usepackage{enumitem}
\usepackage{lscape}
\usepackage{algorithm}
\usepackage{algorithmic}
\usepackage[dvipsnames,table,xcdraw]{xcolor}

\usepackage[numbers]{natbib}
\setcitestyle{authoryear,open={(},close={)}}
\usepackage[breaklinks, colorlinks, citecolor=blue, linkcolor=magenta, urlcolor=blue]{hyperref}
\usepackage[nameinlink]{cleveref}

\newtheorem{theorem}{Theorem}[section]

\newtheorem{rem}[theorem]{Remark}

\newcommand{\cF}{\mathcal{F}}

\newcommand{\cM}{\mathcal{M}}
\newcommand{\cS}{\mathcal{S}}
\newcommand{\cP}{\mathcal{P}}
\newcommand{\N}{\mathbb{N}}
\renewcommand{\P}{\mathbb{P}}
\newcommand{\F}{\mathbb{F}}

\newcommand{\E}{\mathbb{E}}
\newcommand{\R}{\mathbb{R}}
\newcommand{\1}{\mathbf{1}}
\newcommand{\Ereal}{\ensuremath{E_u}}

\newcommand{\piNN}{\ensuremath{\pi^{\text{NN}}}}
\newcommand{\piOracle}{\ensuremath{\pi^{\text{oracle}}}}
\newcommand{\err}{\operatorname{err}}

\let\del\undefined
\let\com\undefined

\if\viewchanges1
	
	\newcommand{\del}[1]{{\color{red}{#1}}}
	\newcommand{\com}[1]{{\color{orange}{#1}}}
	
\else
	
	\newcommand{\del}[1]{}
	\newcommand{\com}[1]{}
	
\fi

\begin{document}

\title{Robust Utility Optimization via a GAN Approach}
\date{\today}

\author{Florian Krach, Josef Teichmann, Hanna Wutte\\
Department of Mathematics, ETH Zürich}

\maketitle



\begin{abstract}
    Robust utility optimization enables an investor to deal with market uncertainty in a structured way, with the goal of maximizing the worst-case outcome.
    In this work, we propose a generative adversarial network (GAN) approach to (approximately) solve robust utility optimization problems in general and realistic settings.
    In particular, we model both the investor and the market by neural networks (NN) and train them in a mini-max zero-sum game.
    This approach is applicable for any continuous utility function and in realistic market settings with trading costs, where only observable information of the market can be used.
    A large empirical study shows the versatile usability of our method. Whenever an optimal reference strategy is available, our method performs on par with it and in the (many) settings without known optimal strategy, our method outperforms all other reference strategies.
    Moreover, we can conclude from our study that the trained path-dependent strategies do not outperform Markovian ones.
    Lastly, we uncover that our generative approach for learning optimal, (non-) robust investments under trading costs generates universally applicable alternatives to well known asymptotic strategies of idealized settings. 
\end{abstract}

\section{Introduction}
In mathematical finance, uncertainty is an inherent and key characteristic
of financial markets and it is essential for risk managers and traders alike to effectively model and
manage uncertainty for making informed financial
decisions.
A pivotal tool for rational decision-making amid this uncertainty is \emph{robust utility optimization}.
Utility optimization enables the balancing act between maximizing returns and managing risk while considering individual preferences and long-term goals. 
To further deal with uncertainty about the law of a given
stochastic processes, in \emph{robust} utility optimization one assumes that there is a family of potential (often parametric) probability laws that the true law belongs to. The focus then lies on building resilient portfolios and strategies that can withstand a wide range of market conditions within these (parametric) laws.

The robust utility maximization problem can be interpreted as a zero-sum game between a trader and an adversarial market -- the \emph{nature}, or more precisely, the uncoordinated accumulation of all other market participants. On the one hand, the trader aims to maximize expected utility of profits in a market environment that is controlled by the nature, who on the other hand, seeks to choose the dynamics of underlying traded assets as unfavorably as possible for the trader. 
Of course, in reality, nature's goal is not to harm a specific trader, however, the advantage of this setting is that it allows for a worst-case analysis. 

The degree to which the antagonistic nature can vary asset dynamics is constrained by a set $\cP$ of potential underlying probabilistic market models. Many variations of constructing such sets have been considered.
For example, the set $\cP$ can be given in parametric terms, or it can be formulated in model-free ways (see, for instance, \citet[Section 2]{bielecki2017adaptive} for a concise overview of existing approaches).
A fairly general approach to constructing potential market measures is to vary drift and diffusion terms in a general SDE formulation. 
The set of measures constructed in that way is vast, and further restrictions on $\cP$ are in order to compute solutions or even guarantee their existence.

One way to (softly) incorporate such constraints is to add penalty terms to the objective of expected utility optimization.
Such an approach was followed by \citet{Guo2022}, who analyze the \emph{penalized robust utility problem} for penalties that penalize deviations of drift and diffusion characteristics from constant reference values.

This approach offers more flexibility beyond fixed interval constraints for drift and diffusion terms.
Nonetheless, the problem remains challenging to solve particularly in realistic market settings. First, explicit solutions even for these simple penalties are known only in standard cases, for log-preferences in friction-less markets \citep{Guo2022}.
Second, classical numerical approaches quickly face limitations as the market dimension increases or trading costs are introduced.

With advances in deep learning (DL) however, new techniques for solving control problems under entirely general constraints have been developed.
In particular, the (constrained) robust portfolio optimization problem is a mini-max game.  For these sorts of problems, generative adversarial network (GAN) approaches \citep{goodfellow2020generative} seem particularly well suited. 
GANs consist of two neural networks (NNs) -- a generator and a discriminator -- that are trained together in a game-like setup. 
As they compete against each other, both networks learn and improve. 
In typical applications, the generator generates new data, while the discriminator tries to distinguish between the generated and real data.
In the context of robust portfolio optimization, the generator outputs portfolio weights and learns to maximize penalized expected utility for a measure $\P$ determined by the discriminator. The discriminator outputs market measures $\P$ with the aim of minimizing that same metric. 
By trading against a malevolent market, the generator is trained to produce NN strategies that withstand a range of adverse scenarios.  

GANs have shown impressive results in various applications, including image and text generation, video synthesis, and drug discovery \citep{GANimage,GANvideo,GANdrugs}.
Through robust approaches, GANs have progressed into mathematical finance as well \citep{cont2023tailgan}.
For robust utility optimization in a one-dimensional market, \cite{Guo2022} test a GAN approach for logarithmic utility functions with robust diffusion. In his simple setting, they find that their GAN approach recovers the  (in this case known) explicit solution. A thorough application of GANs to multidimensional, realistic markets including trading costs in particular is yet outstanding. Moreover, the setup in \citet{Guo2022} contains some unrealistic assumptions on NN architectures that essentially mean the antagonistic market is fully aware of the generator's trades and vice versa. 
Hence there is a lot of room for an in-depth study of GAN technologies in robust portfolio optimization. 
For simplicity, we speak of the \emph{GAN approach} or \emph{GAN strategies} even if we fix the market dynamics and only (non-robustly) train NN approximations of optimal investment strategies.

One major advantage of the GAN approach to (robust) utility optimization is that it is universally applicable across choices of utility functions, market structures, costs and other constraints. In particular, this allows us to apply our method to the problem of optimal investment under transaction costs \citep{constantinides1979multiperiod,davis1990portfolio,de2012optimal,tunc2013optimal,martin2012optimal,martin2016universal,muhlekarbe2017primer,belak2022optimal,sunlearning}, and compare our soultions to well-known \emph{asymptotically} optimal (for vanishing trading costs), non-robust strategies that are characterized by no-trade regions around the optimal friction-less Merton strategy \citep{merton1969lifetime}.

\subsection{Overview of our Contribution}
In this paper, we extend penalized robust utility optimization to realistic, real-world settings. 
In particular, we make the following contributions.
\begin{itemize}
    \item We empirically show that the GAN approaches successfully recover the optimal strategies in multi-asset, friction-less markets with log utilities. 
    \item We further analyze trained strategies for markets with trading costs and general power utilities, where no explicit solution is known.
    \item We introduce fairly general, \emph{path-wise} constraints for adversarial market measures, which allow for more realistic robustification.
    \item Without explicit solutions the results in these more realistic settings require proper evaluation metrics. We propose several metrics for evaluating the learned investment strategies based on a pool of potential market models.
    \item Finally, we compare both \emph{non-robust and robust} GAN-based strategies to investment strategies that are asymptotically optimal for small, vanishing transaction costs in non-robust utility optimization.
    We uncover that non-robustly trained GAN-based strategies differ substantially from the asymptotically optimal ones. Mainly, we observe that contrary to asymptotically optimal strategies, GAN-based strategies
    \begin{itemize}
        \item[(i)] do not incorporate no-trade regions, and
        \item[(ii)] take the time horizon into account  (i.e., they learn when the expected profit over the time horizon is too small compared to the trading costs and act accordingly).
    \end{itemize}
    In spite of these differences, our GAN-based strategies achieve the same expected utility as asymptotically optimal ones while significantly reducing risk.
\end{itemize}

This paper is structured as follows. We first introduce our continuous-time financial market model and the penalized robust utility optimization problem including transaction costs in \Cref{se:Preliminaries}. Then, in \Cref{sec:DeepRobustUtilityOptimization}, we outline our GAN-based algorithm for robust utility optimization in discretized time. In \Cref{sec:Experiments}, we discuss the effectiveness of the GAN-based approach in various experiments: In \Cref{sec:Setting with explicit Solution}, we test the approach in (friction-less) settings with known explicit solutions, in \Cref{sec:MarketWithTransactionCosts} we test it in markets with proportional trading costs, and \Cref{sec:RealisticSettingForStockMarket} contains experiments on robust utility optimization with path-wise penalties. 
Moreover, we further study the effects of small transaction costs in robust and non-robust utility optimization in \Cref{sec:Robust Utility Optimization with Small Trading Cost}. Finally, we conclude our findings in \Cref{sec:Conclusion}.

\section{Preliminaries}\label{se:Preliminaries}
We consider a financial market with finite time horizon $T>0$, consisting of $d\in\N$ risky assets $S=(S^1, \ldots, S^d)$ and a risk-free bank account process $S^0$ compounded at risk-free rate $r\in\R$. 

We work on the Wiener space $\Omega=C([0,T], \mathbb{R}^d)$, with Borel $\sigma$-algebra $\cF$ induced by the topology of uniform convergence on $[0,T]$, Wiener measure $\P$ and as canonical process the $d$-dimensional standard Brownian motion $W$.
Let the dynamics 
of assets $S=(S_{t})_{t\in[0, T]}$ be given as
\begin{equation}\label{eq:assetdynamics}
\begin{split}
    S_0&:=s \in\R^d_{\ge0},\\
        dS_{t} &:= \mu_t\,dt+\sigma_t\,dW_t.
\end{split}
\end{equation}
Here $\mu_t$ and $\sigma_t$ are $\mathcal{F}_t$-measurable random functions taking values in $\R^d$ and $\R^{d\times d}$ respectively, where $\cF_t=\sigma(S_u, u< t)$. The filtration $\F=(\mathcal{F}_t)_{t\in[0, T]}$ captures all market information, progressively in time. 

We further consider an agent trading in this market with strategies that are modeled as portfolio weights $\pi\in\Pi$, where
\begin{align*}
\Pi:=\left\{\pi=(\pi_t)_{t\in[0, T]}: \pi_t\in\R^{d}, \pi \text{ is }\F\text{-predictable}\right\}.\end{align*}
Here, $\pi_t^{i}\in\R^d$ represents the proportion of wealth invested in asset $i$ at time $t_{-}$. The framework allows for short selling ($\pi_t^{i}<0$) and taking up loans ($\sum_{i=1}^d\pi_t^i>1$). The dynamics of the portfolio value process $X^\pi$ with respect to portfolio weights $\pi$ are then given as follows.\footnote{A definition of continuous dynamics for portfolio values including trading costs is given in \cite{muhlekarbe2017primer}. It requires to separately track the portfolio values invested in each asset. For the experiments of this paper, we work in a discrete-time setting where portfolios under transaction costs are defined in \Cref{subsec:DiscreteTimeMarketSetting}. }
We set $X^\pi_{0}:=x_0$, and for $t\ge 0$, 
\begin{align*}
    dX^\pi_{t} &= \sum_{i=1}^d \frac{{X}^\pi_{t}\pi^i_{t}}{S^i_{t}} \,dS^i_{t}+\bigg(1- \sum_{i=1}^d \pi^i_{t}\bigg){X}^\pi_{t}r\,dt.
\end{align*}

In this paper, we assume that the true law $\P^*$ of the asset process $S=(S_{t})_{t\in[0,T]}$ is unknown.
To cope with this uncertainty, we consider the robust portfolio optimization problem 
\begin{align}\label{eq:RobustUO}
    \sup_{\pi\in\Pi}\inf_{\P\in\cP} \E_{\P}\left[u(X_T^\pi)\right],
\end{align}
for a given utility function $u$ and where $\cP$ is a family of possible underlying probabilistic models $\P$.
The family $\cP$ represents epistemic uncertainty. It contains measures $\P$ that generate plausible laws of the underlying assets $S$. In particular, we assume that $\P^*\in\cP$.

We construct potential market measures for $S$ by varying drift and diffusion terms $\mu$ and $\sigma$ within the sets of $\F$-predictable, $\R^d$- and $\R^{d\times d}$-valued processes $\cM$ and $\cS$, respectively. 
Following this approach, we represent laws $\P$ of the underlying assets $S$ in terms of drift and diffusion processes, i.e., we set \begin{align}\label{eq:measureSetwithmuandsigma}
    \cP:=\{\P_{\mu,\sigma}\mid (\mu,\sigma)\in\cM\times\cS\}.
\end{align}

The number of measures constructed in that way is vast, and further restrictions on $\cP$ are in order to arrive at tractable solutions.
Perhaps the most realistic approach to set up potential market measures is to add constraints on $\P\in\cP$ that tie measures $\P$ to observed market behavior
We incorporate such constraints on $\cP$ by adding penalty terms to the objective of \Cref{eq:RobustUO}. More formally, for a penalization functional $R$, let the constrained version of \Cref{eq:RobustUO} be given as
\begin{align}\label{eq:PenalizedRobustUO}
    \sup_{\pi\in\Pi}\inf_{\P_{\mu, \sigma}\in\cP} \E_{\P_{\mu, \sigma}}\left[u(X_T^{\pi, \mu, \sigma})\right]+ R(\P_{\mu, \sigma}),
\end{align}
where the notation $X_T^{\pi, \mu, \sigma}$ makes the dependence of the portfolio value on the choice of the parameters $\mu, \sigma$ defining the market explicit.

\paragraph{Penalties}
A straightforward way to design penalties $R:\cP\to\R_+$ that tie adversarial market measures to observed quantities is to restrict drift and diffusion characteristics to lie within compact balls around reference values inferred from the market at all times. These penalties can be implemented as soft constraints via penalties of the type
\begin{align}\label{eq:GuoPenalty}
R_{\lambda_1, \lambda_2}(\P_{\mu,\sigma})=\E_{\P_{\mu, \sigma}}\left[\lambda_1\int_0^T \lVert \sigma_t\sigma_t^\top  - \tilde  \sigma\tilde\sigma^\top\rVert_F^2\,dt+ \lambda_2 \int_0^T \lVert \mu_t  - \tilde\mu \rVert_2^2\,dt\right],
\end{align}
with Frobenius norm $||\cdot||_F$, and constant reference values for the instantaneous drift $\tilde  \mu\in\R^d$ and covariance $\tilde  \sigma\tilde\sigma^\top \in\R^{d\times d}$, and $\lambda_1, \lambda_2>0$.
Reference values could be obtained either as expert knowledge or via historical estimates.

However, penalties of the type of \Cref{eq:GuoPenalty} face the criticism that reference values, in particular for the drift component $\mu$, can not be estimated well. By extension, these quantities can not be inferred from observing the market, and searching a pool of models tied to these characteristics is not in line with the initial aim.
Consequently, the following question arises. What are adequate market measures to search for in robust portfolio optimization and how do we need to choose the penalties to make this search plausible?

This question is deeply rooted in the field of \emph{stochastic portfolio theory (SPT)} pioneered by \citet{fernholz2009stochastic}.
The central aim of SPT is to develop methods for portfolio optimization that only depend on \emph{observable quantities}.
In SPT, observable quantities are prices, but also properties of paths that can be estimated well. For instance, this includes covariances along an observed trajectory, or some invariant laws observable in the market. 
However, drift coefficients are not considered observable.
Thus, the SPT approach differs from the one of \Cref{eq:GuoPenalty} that ties \emph{both} drift and diffusion terms to certain reference values.

A fairly general way to incorporate observable quantities into the robust utility maximization problem of \Cref{eq:RobustUO} is to include \emph{path-wise} constraints in $\cP$. These constraints can be constructed to relate a potential market measure $\P$ for $S$ to an \emph{observed} trajectory $S^{\mathrm{obs}}$. The idea is to consider any functional $l:C([0,t],\R^{d})\to\R$ with $t \in[0,T]$ and set up a constraint to require that $l(S)$ is close to $l(S^{\mathrm{obs}})$. Here, closeness can for instance be defined via an $L_2$-norm or in $\P$-expectation. 
In terms of soft constraints, these path-wise penalties can be incorporated into \Cref{eq:PenalizedRobustUO} via
\begin{align}
    R_1(\P) &= \lambda_1\int_{\Omega}\left\|{l}(S)-{l}(S^{\mathrm{obs}})\right\|_2^2\,d\P , \text{ or} \label{eq:pwp2} \\
    R_2(\P) &=\lambda_2\left\|\E_\P[{l}(S)]-{l}(S^{\mathrm{obs}})\right\|_2^2. \label{eq:pwp1}
\end{align}
In this paper, we investigate the robust utility optimization problem with path-wise constraints building on the functionals of quadratic covariation up to terminal time $T$, i.e.,
$ l_{\text{QCV}}(S) = [S]_T$ (with penalty $R_1$), 
and relative return over the entire time horizon $l_{\text{RR}}(S)=\frac{S_T}{S_0}$ (with penalty $R_2$). In particular, the latter boils down to computing the \emph{average} relative return
$l_{\text{ARR}}(S) = \E_\P\left[l_{\text{RR}}(S)\right]=\E_\P\left[\frac{S_T}{S_0}\right]$
(see also \Cref{sec:Path-dependent Penalties}).

\begin{rem}\label{rem:minmax theorem}
    If the penalty is of the type \eqref{eq:GuoPenalty} where the drift is fixed and not optimized and $u$ is a continuous, increasing and concave utility function on $\R$, then it was shown in \citet[Proposition 2]{Guo2022} that the $\sup \inf$ and $\inf \sup$ formulation of \eqref{eq:PenalizedRobustUO} coincide. 
\end{rem}

\subsection{Related Work}
An instance of the ``penalization approach'' of \Cref{eq:GuoPenalty} is proposed in \citet{Guo2022}. In this framework, explicit solutions are known for some special cases. In particular, \citet{Guo2022} derive exact solutions to the robust optimization problem for logarithmic utilities and $\lambda_2=0$. Moreover in this setting, \cite{Guo2022} are the first to demonstrate that a deep learning method based on GANs can recover the robust optimal strategy in a one-dimensional market.
Nonetheless, solving a robust utility maximization problem in general markets including frictions, for general utility functions and constraints remains a challenging task. Neither explicit solutions are known, nor do numerical solutions exist so far.
Moreover, to the best of our knowledge, the robust utility optimization problem with path-wise constraints of the form of \Cref{eq:pwp2} or \Cref{eq:pwp1} has not yet been analyzed numerically.

\cite{jaimungal2022robust} introduce another deep learning approach for a different take on the robust portfolio optimization problem \eqref{eq:RobustUO}: Instead of parametrizing measures via drift and diffusion coefficients as we do in \eqref{eq:measureSetwithmuandsigma}, \cite{jaimungal2022robust} characterize the uncertainty set $\cP$ via parametrized perturbations of a truly underlying portfolio distribution. Their set $\cP$ is then further constrained to only contain parameters that cause adversarial perturbations to lie within $\epsilon-$Wasserstein distance to the distribution of the unperturbed portfolio variable. Similarly as in this paper, trading strategies and adversarial market perturbations are then approximated by NNs that are trained on a joint objective. However, an important difference between the approach of \cite{jaimungal2022robust} and ours (and the one of \citet{Guo2022}) is that the setup in \citet{jaimungal2022robust} relies on Markov decision processes, and therefore their deep learning algorithm can not easily uncover path-dependent strategies (or market measures).
Moreover, \cite{jaimungal2022robust} consider more general objectives of rank based expected utility of which expected utility is a special case that they do not treat in their experiments.

\citet{limmer2023robust} use a GAN framework applied to a penalized version of the \emph{robust hedging} problem. In particular, similarly to our work, \citet{limmer2023robust} use two neural networks competing in a penalized zero-sum game, where one models the market and the other one the trading (or hedging) strategy. While our strategy's objective is to maximize the robust expected utility of the terminal wealth, \citet{limmer2023robust} try to minimize the robust distance (measured via a utility functional) between the payoff of a derivative and the wealth achieved with the hedging strategy at terminal time.
Robust hedging is also considered by \citet{wu2023robust} under a similar robustification setting as in \cite{jaimungal2022robust}.

\section{Deep Robust Utility Optimization}\label{sec:DeepRobustUtilityOptimization}
Deep learning has become increasingly popular in mathematical finance for a variety of tasks that are not easily solvable with conventional methods. In the context of robust utility optimization, such an intricate task is finding trading strategies that are robust to adversarial market movements in realistic markets with trading costs. In this section, we first introduce the (realistic) discrete-time market setting for approximating solutions to the penalized robust utility optimization problem \Cref{eq:PenalizedRobustUO} in discrete time. 
Secondly, we propose a GAN-based algorithm for approximating solutions to this problem. This includes approximations of both adversarial market measures and robust trading strategies. 

\subsection{Discrete-time Market Setting}\label{subsec:DiscreteTimeMarketSetting}
For numerically solving the penalized robust utility optimization problem \Cref{eq:PenalizedRobustUO}, we discretize the time interval $[0,T]$ into $N\in\N$ time steps $0=t_0<\ldots<t_N=T$, with step sizes $\Delta t_n := t_n - t_{n-1}$.
Let the discrete-time dynamics 
of assets $S=(S_{t_n})_{n=0,\ldots,N}$ be given by the Euler-Maruyama scheme as
\begin{equation}\label{eq:assetdynamics discrete}
\begin{split}
    S_0&:=s \in\R^d_{\ge0},\\
        S_{t_{n}} &:= S_{t_{n-1}} + \mu_{n-1}\,\Delta t_{{n}}+\sigma_{n-1}\,\Delta W_{t_{n}},
\end{split}
\end{equation}
where $\mu_n$ and $\sigma_n$ are $\mathcal{F}_n$-measurable functions taking values in $\R^d$ and $\R^{d\times d}$ respectively, and $\Delta W_{t_n}$ are increments of a $d$-dimensional standard Brownian motion. 

Furthermore, the discrete-time dynamics with friction 
 of the portfolio value process $X^\pi$ with respect to portfolio weights $\pi$ are given as follows.
We set $X^\pi_{t_0}:=x_0$, and for $n\ge 1$, 
\begin{equation}
    \label{equ:discrete PVP with friction}
\begin{split}
C_{n} &:= (1+r\Delta t_{n+1})\sum_{i=1}^d\Bigg(A_{t_{n}}^iS^i_{t_n}c_{\text{prop}}+c_{\text{base}}\mathrm{1}_{A_{t_{n}}^i>0}\Bigg),\\
    X^\pi_{t_{n+1}} &= {X}^\pi_{t_{n}} + \sum_{i=1}^d \frac{{X}^\pi_{t_{n}}\pi^i_{t_{n}}}{S^i_{t_{{n}}}} \Delta S^i_{t_{n+1}}+\bigg(1- \sum_{i=1}^d \pi^i_{t_{n}}\bigg){X}^\pi_{t_{n}}r\Delta t_{n+1} - C_n,
\end{split}
\end{equation}
with the number of assets $S^i$ traded at time $t_{n-1}$
\begin{align*}
    A_{t_{n}}^i:=\left|\frac{X^\pi_{t_{n}}\pi^i_{t_{n}}}{S^i_{t_{{n}}}}-\frac{X^\pi_{t_{n-1}}\pi^i_{t_{n-1}}}{S^i_{t_{{n-1}}}}\right|.
\end{align*}
Here, we include proportional transaction costs $c_{\text{prop}}$ and additive base costs $c_{\text{base}}$ per trade, with asset dynamics as in \Cref{eq:assetdynamics discrete}. 
Moreover, trading strategies are parametrized as portfolio weights $\pi=(\pi_{t_n})_{n=1,\ldots,N}$, where $\pi_{t_{n}}$ are $\mathcal{F}_n$-measurable functions taking values in $\R^d$.

\subsection{GAN-based Algorithm}
The discrete-time market setting outlined in \Cref{subsec:DiscreteTimeMarketSetting} serves as basis for numerical approaches to approximate saddle point solutions of problem \eqref{eq:PenalizedRobustUO}. In this setting, we look for robust trading strategies  $\pi=(\pi_{t_n})_{n=1,\ldots,N}$ and adversarial market measures $\P_{\mu,\sigma}$ parametrized by $(\mu, \sigma)=(\mu_n, \sigma_n)_{n=0,\ldots,N-1}$, i.e., we search for saddle point solutions to the discrete-time min-max problem
\begin{align}\label{eq:PenalizedRobustUODiscreteTime}
    \sup_{\pi}\inf_{\P_{\mu, \sigma}} \E_{\P_{\mu, \sigma}}\left[u(X_T^{\pi, \mu, \sigma})\right]+ R(\P_{\mu, \sigma}).
\end{align}

Leveraging deep learning in this search, we approximate robust trading strategies $\pi$ and adversarial market characteristics $(\mu, \sigma)$ with NNs $\pi^\theta$ and $(\mu^\phi, \sigma^\phi)$ with parameters $\theta$ and $\phi$, respectively. These NNs are trained to maximize, respectively minimize Monte Carlo estimates of the joint objective
\begin{align}\label{eq:discreteTimeObjective}
\E_{\P_{\mu^\phi, \sigma^\phi}}\left[u(X_T^{\pi^\theta, \mu^\phi, \sigma^\phi})\right]+R(\P_{\mu^\phi, \sigma^\phi}).
\end{align}
Updates of the NNs' parameters $\theta$ and $\phi$ are performed iteratively in a generative adversarial manner with generator $\pi^\theta$ and discriminator $(\mu^\phi, \sigma^\phi)$.

While the order of $\sup$ and $\inf$ in \eqref{eq:PenalizedRobustUODiscreteTime} can be switched in special cases (\Cref{rem:minmax theorem}), this is not necessarily true in general. 
From the perspective of robust utility optimisation, we are interested in the problem as stated in \eqref{eq:PenalizedRobustUODiscreteTime}, i.e., in finding a worst-case optimal trading strategy.
It is important to note that, in general, the GAN learns neither the $\sup \inf$ nor the $\inf \sup$\footnote{For the GAN to learn e.g.\ the $\sup \inf$ of \Cref{eq:PenalizedRobustUODiscreteTime}, the discriminator would need to be trained until it converged to the global minimum after each generator step; and vice versa for learning $\inf \sup$. It can be tried to achieve approximations of these by an imbalance between the number of steps for generator and discriminator.}. Instead it tries to find a regularized\footnote{More precisely, the problem is regularized through the parameterisation of the player's strategies via neural networks and at a stationary point (in the parameter space) the derivative of the objective function with respect to the trainable parameters of each player is $0$.} stationary point, which can be interpreted as a local Nash equilibrium.
However, it is well known that the training of GANs is unstable and usually not even a stationary point is found. But at the same time, the GAN is likely to heavily explore the space of trading strategies. We make use of this by early stopping criteria based on validation metrics approximating the worst-case expected utility. Hence, the trained, early-stopped strategy should approximate the worst-case optimal strategy.

\begin{algorithm}[ht]
    \caption{\texttt{GAN4RUO}}
    \label{alg:GAN}
\begin{algorithmic}[1]
\STATE \textbf{input:} {niter} \COMMENT{number of iterations}, $B$ \COMMENT{batch size}, $\lambda$ \COMMENT{penalty parameter}, $s_0,x_0$ \COMMENT{initial asset and portfolio values}, $D$ \COMMENT{training set of gaussian increments}, $u$ \COMMENT{utility function}
\STATE $\pi^{\theta_0} = \texttt{initialize\textunderscore NN\textunderscore generator}()$
\STATE $\mu^{\phi_0}, \sigma^{\phi_0} = \texttt{initialize\textunderscore NN\textunderscore discriminator}()$
\FOR{$k=0$ to niter-1}
\STATE $D_B =$ \texttt{select\textunderscore B\textunderscore increments}($D, B$)
\STATE $x_T, R_T =$ \texttt{forward}($D_B$, $\pi^{\theta_k},\mu^{\phi_k}, \sigma^{\phi_k}, s_0, x_0$) \COMMENT{compute terminal value and estimate penalties on $D_B$}
    \STATE gen loss $=-\left(\frac{1}{B}\sum u(x_T)\right)-\lambda R_T$
    \STATE disc loss $= -$gen loss
    \IF{\texttt{modulo}(niter, 2) $== 0$} 
    \STATE$\pi^{\theta_{k+1}} = \texttt{train\textunderscore NN}$($\theta_k$, gen loss) 
    \ELSE{    \STATE$\mu^{\phi_{k+1}}, \sigma^{\phi_{k+1}} = \texttt{train\textunderscore NN}$($\phi_k$, disc loss)
}
    \ENDIF

\ENDFOR
\end{algorithmic}
\end{algorithm}

\begin{algorithm}[ht]
    \caption{\texttt{forward}}
    \label{alg:GAN_forward}
\begin{algorithmic}[1]
\STATE \textbf{input:} $D_B$, $\pi^{\theta}, \mu^{\phi}, \sigma^\phi s_0, x_0$
\STATE asset\textunderscore paths $ = []$
\STATE market\textunderscore measures $= []$
\FOR{$t= 0$ to $T-1$}
    \STATE obtain action $a_t = \pi^\theta(t, s_t, x_t)$
    \STATE obtain market parameters $\mu_t = s_t \odot \mu^\phi(t, s_t, x_t)$, $\sigma_t =  s_t \odot \sigma^\phi(t, s_t, x_t)$
    \STATE $s_{t+1}, x_{t+1} = $ \texttt{step}$(s_t, x_t, a_t, \mu_t, \sigma_t, D_B)$
    \STATE asset\textunderscore paths $=$ asset\textunderscore paths $\cup\, s_{t+1}$ \COMMENT{collect asset values}
    \STATE market\textunderscore measures $=$ market\textunderscore measures $\cup\, \mu_t, \sigma_t$ \COMMENT{collect market measures}
\ENDFOR
\STATE $R_T = $ \texttt{estimate\textunderscore path\textunderscore functional}(asset\textunderscore paths, market\textunderscore measures) 
\STATE \textbf{return:} $x_T, R_T$
\end{algorithmic}
\end{algorithm}

An outline of a GAN algorithm (without early stopping criterion) for robust utility optimization is given in \Cref{alg:GAN,alg:GAN_forward}. Following initializations of generator and discriminator networks in lines 2-3 of \Cref{alg:GAN}, these networks are trained on a pre-generated training set of Brownian increments $D$ in lines 4-13 of \Cref{alg:GAN}. For each iteration of the training, a batch of Brownian increments is selected from the training data in line 5 of \Cref{alg:GAN}. Then, in a forward pass through generator and discriminator networks in code line 6, the terminal portfolio value as well as an estimate for the penalty term $R$ are determined for all increments in the current batch $D_B$.
\Cref{alg:GAN_forward} describes the forward pass in which paths for asset values and market measures are collected by consecutive evaluations of generator and discriminator networks. 
We remark that for our experiments we actually use the special form 
\begin{equation}\label{eq:assetdynamics-experiments}
    dS_{t} := S_t \odot \mu_t \, dt+ S_t \odot \sigma_t \, dW_t,
\end{equation}
of the asset dynamics \eqref{eq:assetdynamics} (where $\odot$ is the Hadamard product), as can be seen in line 6 of \Cref{alg:GAN_forward}.
The \texttt{step} function in line 7 of \Cref{alg:GAN_forward} is given by \eqref{eq:assetdynamics discrete}.
Having obtained realizations of terminal portfolio values and estimates of the penalty term based on the training increments $D_B$ in the forward pass, losses for generator and discriminator networks are calculated in lines 7-8 of \Cref{alg:GAN} as Monte Carlo estimates of the (negative) objective \Cref{eq:discreteTimeObjective}. Finally, the NNs are updated in code lines 9-13 alternately over iterations.
Naturally, there are various specifics to be set for these updates such as optimization algorithms and regularization methods. We refer to \Cref{sec:Experiments}
for details on these choices in our experiments.

Moreover, throughout all experiments in \Cref{sec:Experiments}, we put particular focus on uncovering potential path-dependencies in the optimized robust trading strategies (and the optimized characteristics of the adversarial market measure). To that end, we contrast recurrent and feed forward NN architectures for both discriminator and generator networks $(\mu^\phi, \sigma^\phi)$ and $\pi^\theta$, respectively.

\section{Experiments}\label{sec:Experiments}

\paragraph{GAN Setup} We use a similar setup as in \citet{Guo2022}. In particular, we use a GAN approach to train an actor (the generator) and an adversarial market (the discriminator) to approximately solve the penalized robust utility optimization problem. In contrast to their work, both the actor and the market network only get the current stock prices as well as the current portfolio value as input, but not the outputs of each other. This makes the setup much more realistic, since in practice it is neither the case that the actor knows the current instantaneous volatility of the market, nor that the market (which is the collection of all other market participants) knows the current portfolio weightings of the actor. 
Moreover, we propose to use recurrent neural networks (RNNs) instead of different NNs for each time step, which use only the current state as input. This is a more efficient way to allow for time-dependence, but also allows for path-dependence.\\

We start with experiments in easy settings, where an explicit solution exists in \Cref{sec:Setting with explicit Solution}. 
Then, in \Cref{sec:MarketWithTransactionCosts,sec:RealisticSettingForStockMarket}, we introduce friction, different utility functions and more realistic path-dependent penalization. In all of these cases, explicit solutions are not known.
Therefore, in \Cref{sec:Evaluation Criterion: Early Stopping and General Assessment of Robustness}, we introduce different metrics to evaluate the performance of our model. 
Moreover, in \Cref{sec:Robust Utility Optimization with Small Trading Cost}, we leverage our GAN-based algorithm in an experimental study of small, proportional trading costs on optimal investments in robust and non-robust utility optimization.

The code with all experiments is available at \url{https://github.com/FlorianKrach/RobustUtilityOptimizationGAN}. Since we compare results in many different settings, we mainly do not show raw results here (those can be obtained by running the respective code), but instead give qualitative explanations of the results.

\subsection{Friction-less Settings with Explicit Solutions}\label{sec:Setting with explicit Solution}

In the following examples we always use stock markets with risk free interest rate $r=1.5\%$, step size $\delta_t = \frac{1}{65}$ and $65$ steps. All stocks start at the initial value $S_0=1$ and are given by an It\^o-diffusion driven by one or multiple Brownian motions. We sample a total of $200$K paths of Brownian motions, which are used for all experiments in this section such that there is no random influence due to different samples across different algorithm evaluations, with an $80/20$ train/test split.\\

We use the explicit solution for the market volatility $\Sigma^*$ (and $\mu^*$ if the drift is also robustified) to get an evaluation metric that computes the expected utility of either the learned or the explicit investment strategy against the explicit solution for the market. Let us define
\begin{equation}\label{eq:explicitSigExpectedUtility}
    E_u(\pi, \Sigma, \mu) := \E\left[ u(X_T^{\pi, \mu, \Sigma}) \right],
\end{equation}
then we compute $E_u(\pi, \Sigma^*, \mu^*)$
for the learned strategy $\pi=\piNN$ and for the explicit strategy $\pi=\pi^*$ (where $\mu^*=\mu$ in case the drift is fixed) and compare them via their relative error
\begin{equation*}
    \err_{rel}^{\Sigma^\star, \mu^*}(\piNN, \pi^\star) := \frac{E_U(\pi^\star, \Sigma^\star, \mu^*) - E_U(\pi_\theta, \Sigma^\star, \mu^*) }{| E_U(\pi^\star, \Sigma^\star, \mu^*) | } \,.
\end{equation*}
We note that $\err_{rel}^{\Sigma^\star, \mu^*}(\piNN, \pi^\star) < 0$ means that $\piNN$ performs better than $\pi^*$.

\subsubsection{Stock Market with 1 Risky Asset \& Robust Volatility}\label{sec:Stock Market with 1 Risky Asset}
As a sanity check, we start with the experimental setting of the GAN example in \citet{Guo2022}.
As in \citet{Guo2022} we use a setting where the drift is fixed for all stocks to be $\mu=3.5\%$ and only the volatility is robustified and, therefore, given by the market's neural network output.
The utility function is $u(x) = \ln(x)$.
As penalty we use 
\begin{equation*}
    R_{\lambda_1}(\P_{\mu,\sigma})=\E_{\P_{\mu, \sigma}}\left[\lambda_1\int_0^T(\sigma_t - \tilde \sigma)^2\,dt\right],
\end{equation*} where $\tilde \sigma\in\R_+$ is a constant reference value for the volatility (this could for example be estimated from historical data), and $\lambda_1$ is the scaling factor for the penalization.
In this case there exists an explicit solution for the optimal actor and market strategy, which was given in \citet[Section 7.1]{Guo2022}. In particular, the optimal investment strategy and volatility are both constant and given as the solution of
\begin{equation*}
    \pi = \frac{\mu -r}{ \sigma^2}, \quad 0= \sigma^4 - \tilde \sigma  \sigma^3 - \frac{(\mu - r)^2}{2 \lambda_1},
\end{equation*}
which we solve numerically to get a reference value.
We use this setting to test and compare different models, since it offers an easy evaluation metric. 
In particular, we compute the expected utility (without penalty) when using the actor's trained investment strategy and the explicit solution for the volatility (evaluation metric) and compare it to the expected utility when using the explicit solution for the actor and the market (reference evaluation metric).
The reason for this type of evaluation procedure is that our ultimate goal is to achieve a good robust investment strategy from the GAN training, which should perform well in the harshest possible market environment.

As in \citet{Guo2022}, we used as learning rate $5\cdot 10^{-4}$, which we decreased by a factor of $0.2$ every $100$ epochs; we trained for $150$ epochs in total with the Adam optimizer with the default hyper-parameters $\beta = (0.9, 0.999)$. 
The initial wealth was set to $X_0 = 5$, the penalty scaling factor to $\lambda_1 = 10$ and the reference value $\tilde\sigma = 0.25$.
We tested different batch sizes $B \in \{ 100, 1000, 10'000 \}$  of which $1000$ turned out to work best. Moreover we tested different relations of number of steps of the discriminator and generator  (5-1, 1-1, 1-5) and different network architectures ($1$-layer feed-forward network, $1$-layer RNN, $4$-layer time-grid network, i.e. for every time step a different NN is used), however, all of them led to similarly good results in this easy setting, with $\err_{rel}^{\Sigma^\star, \mu}(\piNN, \pi^\star) < 0.2\%$.

\subsubsection{Stock Market with Multiple Risky Assets \& Robust Volatility}\label{sec:StockMarketwithMultipleRiskyAssets}
In all experiments of this section, we consider $d \in \N$ risky assets that evolve with the same fixed drift $\mu = \1_d \cdot 3.5\%$ (i.e., we only robustify against the volatility) and start from the same initial value $S_0= \1_d$. The reference value for the volatility $\tilde \sigma \in \R^{d \times d} $ is the matrix square root of the reference covariance matrix $\tilde \Sigma = \tilde \sigma \tilde \sigma^\top$.
As penalty function we  use 
\begin{equation}\label{eq:additive instant penalty}
    R_{\lambda_1}(\P_{\mu,\sigma})=\E_{\P_{\mu, \sigma}}\left[\lambda_1\int_0^T\lVert \Sigma_t  - \tilde  \Sigma \rVert_F^2\,dt\right],
\end{equation} 
where $\Sigma_t = \sigma_t \sigma_t^\top$. The utility function is $u(x) = \ln(x)$.
To derive the explicit solution of the (continuous-time) optimization problem \eqref{eq:PenalizedRobustUO}, we first use It\^o's lemma to derive the continuous dynamics of $\ln(X_T^{\pi, \mu, \sigma})$. After dropping the martingale terms, \eqref{eq:PenalizedRobustUO} simplifies to
\begin{equation}\label{eq:MultiDObjective}
    \sup_{\pi} \inf_{\Sigma} \E^{t,x}\left[\ln(x) + \int_t^T \pi_s^\top (\mu - r) + r - \frac{1}{2} \pi_s^\top \Sigma_s \pi_s + \lambda_1 \lVert \Sigma_s  - \tilde  \Sigma \rVert_F^2\, ds \right].
\end{equation}
We follow \cite[Section 7.2]{Guo2022} and compute the instantaneous derivatives of the integrand with respect to $\pi_t$ and $\Sigma_t$ and set them to $0$. This leads (after some rearrangement) to the equation systems
\begin{equation}\label{eq:explicit system 1}
    \Sigma = \frac{\pi \pi^\top}{4 \lambda_1} + \tilde \Sigma \quad \text{and} \quad \Sigma \pi = (\mu - r).
\end{equation}
The explicit solutions $\pi^*, \Sigma^*$ can then be computed  by first substituting $\Sigma$ in the second equation and numerically solving it for $\pi$ to obtain $\pi^*$ and then using this to compute $\Sigma^*$.

We analyzed various stock market settings, with

\begin{itemize}
    \item 2 uncorrelated risky assets with penalty $\lambda_1=1$
    \begin{itemize}
        \item[(S)] a symmetric market with reference volatility $\tilde \sigma = \begin{pmatrix}
        0.25 & 0 \\ 
        0 & 0.25
    \end{pmatrix},$
    and \label{item:S}
    \item[(AS)] an asymmetric market
    with reference volatility $\tilde \sigma = \begin{pmatrix}
        0.15 & 0 \\ 
        0 & 0.35
    \end{pmatrix}.$\label{item:AS}
    \end{itemize}

    \item 2 correlated risky assets with penalty $\lambda_1=1$
    \begin{itemize}
        \item[(PS)] positively correlated, symmetric setting: both assets' volatilities are $0.25$, their correlation is $0.9$, leading to the reference volatility 
        \begin{align*}
            \tilde \sigma = \begin{pmatrix}
        0.25 &  0\\
        0.225 & 0.10897247
    \end{pmatrix},
        \end{align*} \label{item:PS}
        \item[(PAS)] positively correlated, asymmetric setting: assets' volatilities are $0.15$ and $0.35$, their correlation is $0.9$, leading to the reference volatility 
        \begin{align*}
            \tilde \sigma = \begin{pmatrix}
        0.15 &  0\\
        0.315 & 0.15256146
    \end{pmatrix},
        \end{align*} \label{item:PAS}
        \item[(NAS)] negatively correlated, asymmetric setting: assets' volatilities are $0.15$ and $0.35$, their correlation is $-0.9$, leading to the reference volatility 
        \begin{align*}
            \tilde \sigma = \begin{pmatrix}
        0.15 &  0\\
        -0.315 & 0.15256146
    \end{pmatrix}.
        \end{align*} \label{item:NAS}
    \end{itemize}

    \item 5 uncorrelated risky assets with penalty $\lambda_1 \in \{0.1, 1, 10\}$
    \begin{itemize}
        \item[(5S)] uncorrelated, symmetric setting $\tilde \sigma = 0.25\cdot I_5$.
    \end{itemize}
\end{itemize}
In all cases, our model achieves a very small relative error compared to the explicit strategy $\pi^*$  satisfying $\err_{rel}^{\Sigma^\star, \mu}(\piNN, \pi^\star) < 0.2\%$.

\subsubsection{Robustifying Volatility and Drift}\label{sec:Robustifying Volatility and Drift}

In this section, we robustify against uncertainty in the drift in addition to uncertainty in the volatility. Analogously to the robustification w.r.t.\ the volatility of \Cref{sec:StockMarketwithMultipleRiskyAssets}, we choose a reference value $\tilde{\mu}$ for the drift and consider the joint penalty function \eqref{eq:GuoPenalty}.
Then, for log-utility $u = \ln$, we can compute  the explicit solution as in Section~\ref{sec:StockMarketwithMultipleRiskyAssets}.
In particular, we first apply It\^o's lemma to \eqref{eq:PenalizedRobustUO}, then drop the martingale terms and compute the instantaneous derivative of the integrand with respect to $\pi_t, \Sigma_t$ and $\mu_t$. Setting them to $0$ yields
\begin{equation*}
    \Sigma = \frac{\pi \pi^\top}{4 \lambda_1} +  \tilde\Sigma \quad \text{and} \quad \Sigma \pi = (\mu - r) \quad \text{and} \quad \pi = 2\lambda_2 (\tilde\mu - \mu),
\end{equation*}
which can be solved numerically for the unique explicit solution $\pi^*, \Sigma^*, \mu^*$.
Therefore, we can again compute the relative errors $\err_{rel}^{\Sigma^\star, \mu^*}(\piNN, \pi^\star)$.\\

We test our GAN approach in a stock market with $d=2$ risky assets. For $\Sigma$, we use the settings (S), (AS), (PS), (PAS) and (NAS) defined in Section~\ref{sec:StockMarketwithMultipleRiskyAssets}. Moreover, for the drift we test
\begin{itemize}
    \item[(SD)] a symmetric reference drift
    \begin{align*}
    \tilde{\mu} = \begin{pmatrix}
        0.035\\
        0.035
    \end{pmatrix},
    \end{align*} \label{item:SD}

    \item[(AD)] and an asymmetric reference drift
    \begin{align*}
    \tilde{\mu} = \begin{pmatrix}
        0.035\\
        0.055
    \end{pmatrix},
    \end{align*} \label{item:AD}
\end{itemize}

and use balanced penalty weights $\lambda_1 = \lambda_2 = 1$ (cf.\ \Cref{eq:GuoPenalty}).
In all combinations, we find a relative error $\err_{rel}^{\Sigma^\star, \mu^*}(\piNN, \pi^\star) < 0.2\%$.

\begin{figure}[!ht]
    \centering    
    \includegraphics[width=.49\textwidth]{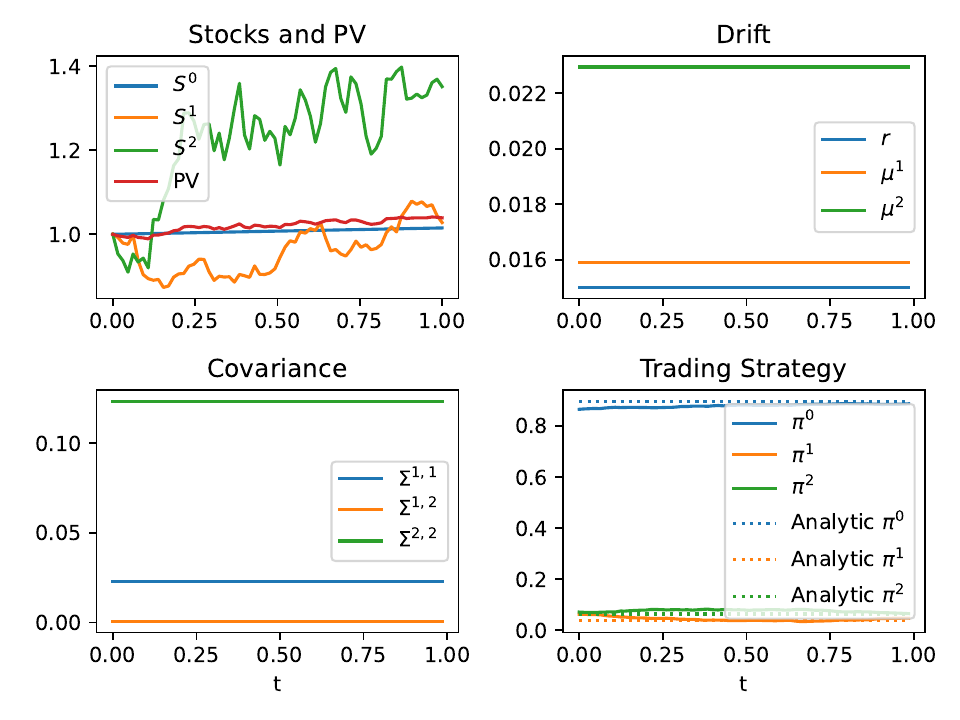}   
    \includegraphics[width=.49\textwidth]{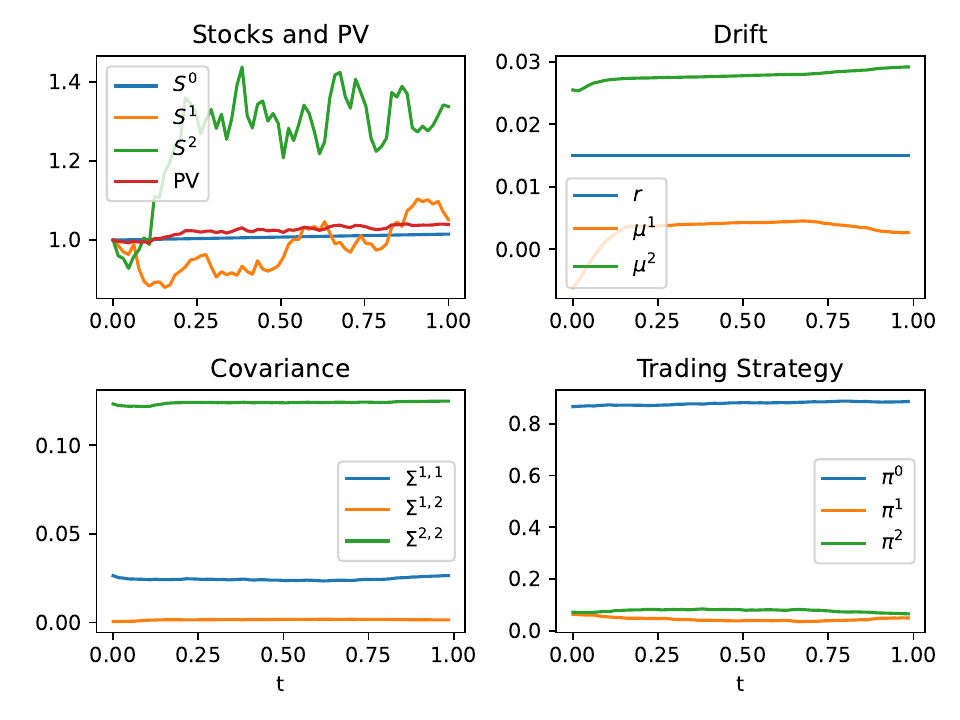} 
    \caption{Test sample paths of the trained trading strategy evaluated against the explicit solution for the market (left) and against the trained discriminator for the market (right). A FFNN is used for generator and discriminator.}
    \label{fig:trading strategy vs explicit and trained market}
\end{figure}

Moreover, we get similar results when replacing the recurrent neural network (RNN) structures of generator and discriminator by standard feed-forward neural networks (FFNN) (that are reused for each time step). This was anticipated, since the optimal explicit strategy is constant and therefore not path dependent. In Figure~\ref{fig:trading strategy vs explicit and trained market}, we show sample paths of the trained trading strategy evaluated against the explicit solution $\Sigma^\star,\mu^\star$ for the market, and against the trained discriminator for the market on the test dataset.
We see that in both cases, the trained strategy resembles the explicit strategy quite well. Interestingly, we further observe that the trained discriminator produces volatilities fairly similar to the explicit solution $\Sigma^\star$ (bottom left panels in \Cref{fig:trading strategy vs explicit and trained market}. This is not the case for adversarially learned drift values that differ from $\mu^\star$ (top right panels in \Cref{fig:trading strategy vs explicit and trained market}).

\subsection{Settings with Proportional Transaction Costs}\label{sec:MarketWithTransactionCosts}
We consider the discrete portfolio value process~\eqref{equ:discrete PVP with friction} with  proportional transaction costs $c_{\text{prop}}$ and additive base costs $c_{\text{base}}$ per trade. 
We stick to the market settings of Section~\ref{sec:Robustifying Volatility and Drift} and, as before, apply our algorithm to approximate solutions to~\eqref{eq:RobustUO} with log-utility $u=\ln$.
We consider market settings with low and high proportional cost $c_{\text{prop}} \in \{ 0.1\%, 1\% \}$ respectively. In all experiments, we set $c_{\text{base}}:=0$. 

Since explicit solutions $\pi_c^*,\Sigma_c^*$ (and $\mu^*$) for the problem including transaction costs are not known, we instead use the explicit solution for the corresponding problem without transaction costs (problem \eqref{eq:MultiDObjective}), which can be considered as the best strategy that can be derived explicitly. 
In contrast to this limitation of closed-form solutions, a simple adjustment of the training framework allows us to train our GAN model in the market with transaction costs.
As before, we compute the relative errors $\err_{rel}^{\Sigma^\star, \mu^*}(\piNN, \pi^\star)$, where market friction is in place. 
Clearly, the explicit market parameters are, in general, not the optimizers of the problem with friction, however, since our goal is to achieve a robust strategy, it is still reasonable to test the strategies against these (maybe suboptimal) market parameters.
We note again that negative values correspond to superior performance of the GAN-based method.

We consider the same settings for the reference volatility $\tilde{\Sigma}$ and reference drift $\tilde{\mu}$ as in \Cref{sec:Robustifying Volatility and Drift}.
For small transaction costs $c_{\text{prop}}= 0.1\%$, we get $\err_{rel}^{\Sigma^\star, \mu^*}(\piNN, \pi^\star) \in (-2.3\%, -0.3\%)$, which indicates slight outperformance of the (friction-aware) GAN model compared to the explicit solution $\pi^\star$ of the friction-less case. In particular, we see that taking friction into account improves the performance of the strategy, but that the difference is not very large due to the small size of the transaction costs.
For larger transaction costs $c_{\text{prop}}= 1\%$, this effect becomes considerably stronger, with relative errors resulting in $\err_{rel}^{\Sigma^\star, \mu^*}(\piNN, \pi^\star) \in (-165.3\%, -10.5\%)$.

Again, we did not see a different performance when using FFNNs performed on par with RNNs, suggesting that the optimal strategy is not path-dependent, even in the market with friction. We note, however, that this empirical result is no (empirical) proof for no path-dependence, since the optimal strategy might outperform both, the trained FFNN and RNN strategies.

\subsection{Realistic Settings for Generic Penalized Robust Utility Optimization}\label{sec:RealisticSettingForStockMarket}
In this section, we analyze what we consider the most realistic setting for penalized robust utility optimization: We robustify against uncertainty in both the volatility and the drift, in a market with transaction costs, where we consider general power utilities corresponding to various preference structures including but not limited to log-preferences. Moreover, we study the path-wise penalties $R_1$ and $R_2$ of \Cref{eq:pwp1,eq:pwp2}. These penalties are more realistic than the penalties of \Cref{sec:Setting with explicit Solution,sec:MarketWithTransactionCosts}, since $R_1$ and $R_2$ can be estimated well from observed stock paths in contrast to the (empirically) unobservable instantaneous volatility and drift parameters that appear in \Cref{eq:additive instant penalty}.

In this general setting, no explicit solution to the robust utility optimization problem is known.
Moreover, this setting differs in too many regards from the setting in \Cref{sec:Setting with explicit Solution}, where only log utility, no transaction costs and instantaneous penalization were considered, such that it is not reasonable to benchmark against the explicit solutions of \Cref{sec:Setting with explicit Solution} anymore. Therefore, we introduce new metrics to evaluate model performance in \Cref{sec:Evaluation Criterion: Early Stopping and General Assessment of Robustness}. Prior to that, we define general power utility functions and realistic path-wise penalties in \Cref{sec:Power Utility as Generalization of Log Utility,sec:Path-dependent Penalties}.

\subsubsection{Power Utility as Generalization of Log Utility}\label{sec:Power Utility as Generalization of Log Utility}
\begin{figure}[!tb]
    \centering    \includegraphics[width=.6\textwidth]{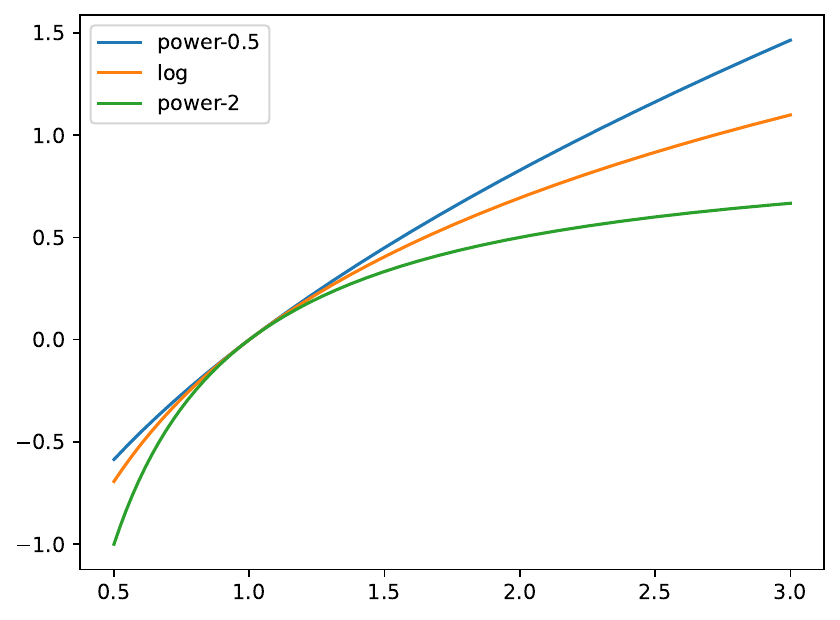}    
    \caption{Utility functions $u_p$ for different powers $p$.}
    \label{fig:utility functions}
\end{figure}

We consider the power utility functions
\begin{equation}\label{eq:ourpowerutility}
    u_p(x) = \frac{x^{1-p} -1}{1-p},
\end{equation}
where $p \in \R_{\geq 0}\setminus\{ 1\}$. In the limit $p \to 1$ the $\log$-utility function is recovered, hence we write $u_1 := \ln$, and for $p=0$ $u_p$ equals the risk-neutral utility $id -1$. In particular, $p\in (0,1)$ leads to less risk aversion compared to the $\log$-utility function, while $p > 1$ leads to more risk aversion.
In the following, we use $p \in \{ 1, \frac{1}{2}, 2 \}$, corresponding to the log utility, a more risk-seeking, and a more risk-averse option, respectively. The graphs of these utility functions are plotted in \Cref{fig:utility functions}.

\subsubsection{Path-dependent Penalties}
\label{sec:Path-dependent Penalties}
In this section, we analyze our GAN-based approach for the penalized robust utility optimization problem of \Cref{eq:PenalizedRobustUO} under the general path-dependent penalties of \Cref{eq:pwp1,eq:pwp2}. 
In contrast to the penalties of \Cref{eq:GuoPenalty} that we considered in \Cref{sec:Setting with explicit Solution,sec:MarketWithTransactionCosts}, these path-dependent penalties can be estimated well from observable quantities.

Moreover, observing the paths of trained generator and discriminator outputs $\piNN$ and $\mu^\theta$, $\Sigma^\theta$, in \Cref{fig:trading strategy vs explicit and trained market} respectively, we notice the following: in the setting of \Cref{sec:Robustifying Volatility and Drift} which included point-wise penalties as in \Cref{eq:GuoPenalty}, the adversarial has learned to produce adversarial market parameters that almost stay constant, even in the fully robust case, where the adversarial was allowed to vary both drift and diffusion terms.
This leads us to believe that the penalty \eqref{eq:GuoPenalty} that was suggested in \citet{Guo2022} and which we have considered so far, was too strict. 
The penalty  required the adversarial instantaneous drift and diffusion coefficients $\mu^\theta_t$ and $\Sigma^\theta_t$ to be close to $\tilde\mu$ and $\tilde\Sigma$, respectively,  \emph{at all times} during the training. Indeed, such penalties might not allow for much exploration.

Instead, in this section we consider \emph{path-wise} penalties $R_1, R_2$ that build on an observed data path, as introduced in \Cref{eq:pwp1,eq:pwp2}. 
In particular, we consider the quadratic covariation $\ell_{QCV}$ penalized via \eqref{eq:pwp2} (which is a constraint for the volatility) and the average relative return $l_{\text{ARR}}$ penalized via \eqref{eq:pwp1} (which is a constraint for the drift).
Since we work with discretized processes on a time grid $0=t_0<\ldots<t_N=T$, we approximate the quadratic covariation and average relative return functionals $\ell_{QCV}$ and $l_{\text{ARR}}$, with 
\begin{align}
    \hat l_{\text{QCV}}(S)_{i,j}&:=\sum_{n=1}^N(S_{t_n}^i-S_{t_{n-1}}^i)(S_{t_n}^j-S_{t_{n-1}}^j), \label{eq:QuadCovarApprox}\\
    \hat l_{\text{ARR}}(S)_i&:=\frac{1}{|B|}\sum_{b\in B}S_T^i(\omega_b)/S_0^i(\omega_b), \label{eq:AvRelRetApprox}
\end{align}
respectively, where $B$ is a batch of paths generated by the adversarial market.

In synthetic settings, where we do not have observed data as reference, we assume a reference market \eqref{eq:assetdynamics-experiments} with constant parameters $\tilde\mu, \tilde\Sigma$ and replace $S^{\mathrm{obs}}$ by $S^{\tilde\mu, \tilde\Sigma}$ in \Cref{eq:pwp2,eq:pwp1}.
In particular, this corresponds to a Black--Scholes model as reference market.
The corresponding reference values are
\begin{align*}
    l_{\text{QCV}}(\ln S^{\tilde\mu, \tilde\Sigma})=\tilde\Sigma\cdot T,\quad   l_{\textrm{ARR}}(S^{\tilde\mu, \tilde\Sigma})=e^{\tilde\mu\cdot T},
\end{align*}
where we compute the quadratic covariation of log-assets. 
In line with this, we also compute $\hat l_{\text{QCV}}(\ln S)$, i.e., the quadratic covariation of the logarithm of the paths of the adversarial market.

\subsubsection{Evaluation Criterion: Early Stopping and General Assessment of Robustness}\label{sec:Evaluation Criterion: Early Stopping and General Assessment of Robustness}
\paragraph{Assessment of Robustness.}
Since no explicit solution is known, we have to define new metrics to asses the robustness and quality of trained models, and to perform early stopping during training. We propose the following metrics, which are all based on a reference market model. In real world settings, this reference market model is obtained by selecting a period of historical test data on which the reference parameters $\tilde \Sigma,\tilde\mu$ and their respective (co)variance are estimated. 
For synthetic experiments, we assume that reference parameters $\tilde\Sigma,\tilde\mu$ and their (co)variance are given. 

Based on the reference model, we use one of the following approaches to generate a pool of market scenarios $S^j$ for $1\leq j \leq n_{\text{pool}}$, where each $S^j$ should be understood as a stochastic process (rather than one realization of it) that corresponds to one market measure $\P_j \in \mathcal{P}$.
\begin{enumerate}
    \item \textbf{Noisy Market Models around Reference Market.}
    We generate random noisy market scenarios $\Sigma^{\mathrm{noisy}, j}_t,\mu^{\mathrm{noisy}, j}_t$ in one of the following 3 ways.
    \begin{itemize}
        \item (Constant noise) For each $j$ we sample one realization from the normal distributions with means $\tilde\Sigma$ and $\tilde\mu$, and their corresponding (co)variances and set $\Sigma^{\mathrm{noisy}, j}_t,\mu^{\mathrm{noisy}, j}_t$ equal to these samples for all time points $t$.
        \item (Non-constant noise) For each $j$ and $t$ we sample one realization from the normal distributions with means $\tilde\Sigma $ and $\tilde\mu$ and their corresponding (co)variances and set $\Sigma^{\mathrm{noisy}, j}_t,\mu^{\mathrm{noisy}, j}_t$ equal to these samples (i.e. noisy around $\tilde\Sigma,\tilde\mu$).
        \item (Cumulative noise) For each $j$ we sample a Brownian motion path starting at $\tilde\Sigma,\tilde\mu$ and scaled by their (co)variances divided by the square-root of the number of steps (s.t. the last step has the given (co)variance). We set $\Sigma^{\mathrm{noisy}, j}_t,\mu^{\mathrm{noisy}, j}_t$ equal to these paths (i.e. noisy path starting at $\tilde\Sigma,\tilde\mu$).
    \end{itemize}
    We remark that the cumulative noise type is the most realistic one for real world market behaviour, since parameter uncertainty increases with time. In the subsequent evaluations of \Cref{sec:Experimental Results}, we therefore focus on cumulative noise. In this setting, the market scenario $S^j$ is thus given by the SDE \eqref{eq:assetdynamics-experiments} with parameters $\Sigma^{\mathrm{noisy}, j}_t,\mu^{\mathrm{noisy}, j}_t$ (where the Cholesky decomposition is used to compute a standard deviation matrix $\sigma$ from a variance matrix $\Sigma$).

    \item \textbf{Noisy GARCH Models around Reference Market.}
    In an alternative approach, we fit a multidimensional GARCH model to the historical test data (or paths independently generated from the synthetic market model with the reference parameters $\tilde\Sigma,\tilde\mu$). The multivariate fit is achieved by first fitting a GARCH model for each coordinate and then computing the correlation matrix between the standardized residuals, which is used to get a joint distribution of all coordinates. Then, we construct a pool of noisy models around this fitted GARCH model, by adding Gaussian noise to the fitted GARCH parameters and to the correlations between the coordinates. Each noisy GARCH model corresponds to one market scenario $S^j$.
\end{enumerate}
For each $S^j$ of the pool of market scenarios, we then approximate the expected utility 
\begin{equation*}
    E_u(\pi, S^j) := \E\left[ u(X_T^{\pi, S^j}) \right],
\end{equation*} 
where $X^{\pi, S}$ is the wealth process when applying strategy $\pi$ in the market with stock path $S$. In particular, for each fixed $\Sigma^{\mathrm{noisy}, j}_t,\mu^{\mathrm{noisy}, j}_t$ or for each fixed noisy GARCH parameters, we sample a batch of $B_{\text{test}}$ i.i.d.\ paths and compute a Monte Carlo estimate to approximate the expectation.
For each strategy, the metric $M_u$ is then defined as the minimum across these (approximate) expected utilities, 
\begin{equation}\label{equ:eval metric}
    M_u(\pi) := \min_{1 \leq j \leq n_{\text{pool}}} E_u(\pi, S^j).
\end{equation}
In particular, $M_u$ reports the worst case expected utility among the pool of reasonable market scenarios. This is an approximation of 
\[ \inf_{\P_{j}\in\cP} \E_{\P_{j}}\left[u(X_T^{\pi, S^j})\right],\] 
which is the quantity we would like to maximise according to \eqref{eq:PenalizedRobustUO}.
We use this metric $M_u$ for evaluating robustness of competing strategies on a test data set, where $E_u$ is approximated with $B_{\text{test}}$ test samples.

\paragraph{Early Stopping.}
Ideally, we would apply metric $M_u$ for model selection in the GAN training of $\piNN$ as well. However, computing a large enough number of expected utilities at each epoch is computationally costly. Instead of $M_u$, we therefore use the average over expected utilities, 
\begin{equation}\label{equ:early stopping metric}
    \frac{1}{B_{\text{val}}}\sum_{j=1}^{B_{\text{val}}} E_u(\pi, S^j)
\end{equation}
for model selection, where for each expected utility a one-sample Monte Carlo approximation is computed from the corresponding sample of the validation batch. 
Since the noise samples to get the market scenarios $S^j$ are independent of the increment samples used to produce paths of $S^j$, Fubini's theorem implies that this approximation of \eqref{equ:early stopping metric} converges to 
\[ \E_{\P_j \in \mathcal{P}}\left[ \E_{S^j \sim \P_j}\left[ u(X_T^{\pi, S^j}) \right] \right], \]
by the law of large numbers. Hence, choosing $B_{\text{val}}$ large enough suffices for this to yield a reliable approximation of this metric, which is also efficient-to-compute. 
In contrast to this, using one-sample approximations for \eqref{equ:eval metric} would yield a bad approximation of the metric, since it uses the minimum instead of the average, which means that the law of large numbers would not apply.
Since \eqref{equ:early stopping metric} is clearly less pessimistic than \eqref{equ:eval metric}, we suggest to counteract by choosing higher noise levels for computing \eqref{equ:early stopping metric} in comparison to the (expected) noise levels of the evaluation metric \eqref{equ:eval metric}.

\paragraph{Evaluation on Reference Market Model.}
Another sanity check for the robustly trained strategies is to compare them to the non-robust models (i.e., the model where only the generator is trained, while the discriminator is fixed to the reference parameters $\tilde\Sigma,\tilde\mu$) when both are applied to a market following the reference parameters $\tilde\Sigma,\tilde\mu$. In particular, this measures the cost of robustification, given that the market is not deviating from the reference market model. 
Clearly, one cannot hope for the same performance of the robust strategy as for the non-robust one in terms of expected utility in the reference market, since there is a price to pay for the additional robustness.
Therefore, we propose to compare the distributions of the terminal wealth $X_T^{\pi}$ for the different strategies $\pi$, because this allows for more insight into the effect of the robustification.

\subsubsection{Experimental Results}\label{sec:Experimental Results}
In this section, we report the results of a series of experiments on realistic penalized robust utility optimization as outlined earlier in \Cref{sec:RealisticSettingForStockMarket}.
In these experiments, we train and compare non-robust, robust volatility and fully robust models in a 2-dimensional market with setting (AS) for the reference volatility $\tilde \Sigma$, setting (AD) for the reference drift $\tilde \mu$, and all of interest rate, step size, and initial asset value as in \Cref{sec:Setting with explicit Solution}.
We include proportional transaction costs $c_{\text{prop}} = 1\% $, consider the three utility functions $u \in \{u_{1/2} , u_1=\ln, u_{2} \}$ and path-dependent penalties as defined in \Cref{sec:Path-dependent Penalties}.

\paragraph{Evaluation with Noisy Market Models.}
For the model evaluation on the test set we use the standard deviations of $0.075$ for the volatility and $0.01$ for the drift. To counteract the too optimistic view of the validation metric \eqref{equ:early stopping metric}, we double these standard deviations, i.e., $0.15$ for the volatility and $0.02$ for the drift, for the early stopping validation during training. 
Moreover, we use $n_{\text{pool}}=1000$, $B_{\text{test}} = B_{\text{val}} = 40K$ test and validation samples and $160K$ training samples.
As outlined earlier in \Cref{sec:Evaluation Criterion: Early Stopping and General Assessment of Robustness}, we only report results with the cumulative noise type. 

\begin{figure}[!htbp]
    \centering    
    \includegraphics[width=.45\textwidth]{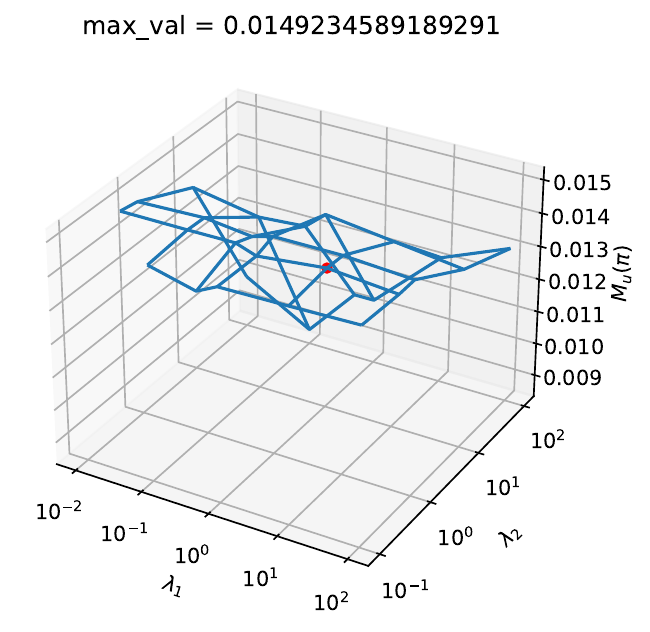} 
    \includegraphics[width=.45\textwidth]{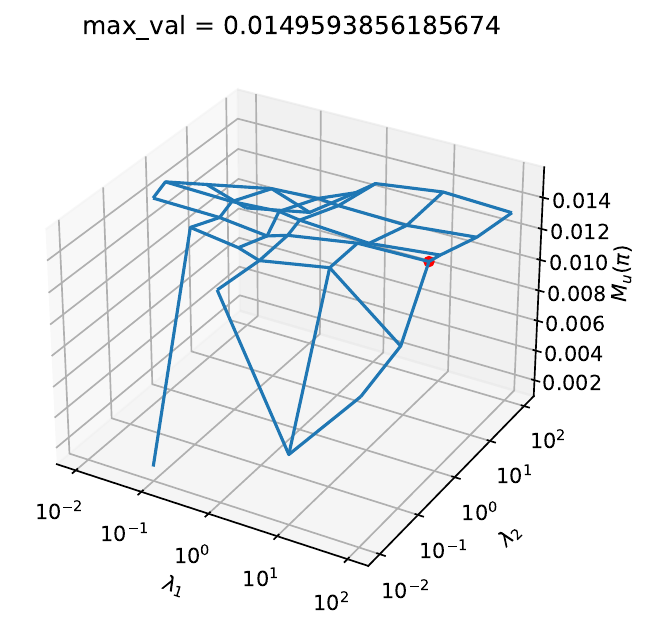} 
    \includegraphics[width=.45\textwidth]{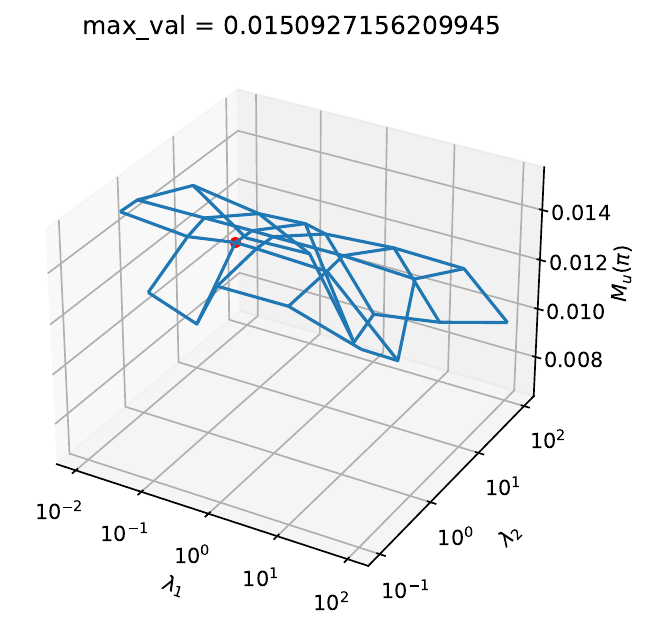}   
    \includegraphics[width=.45\textwidth]{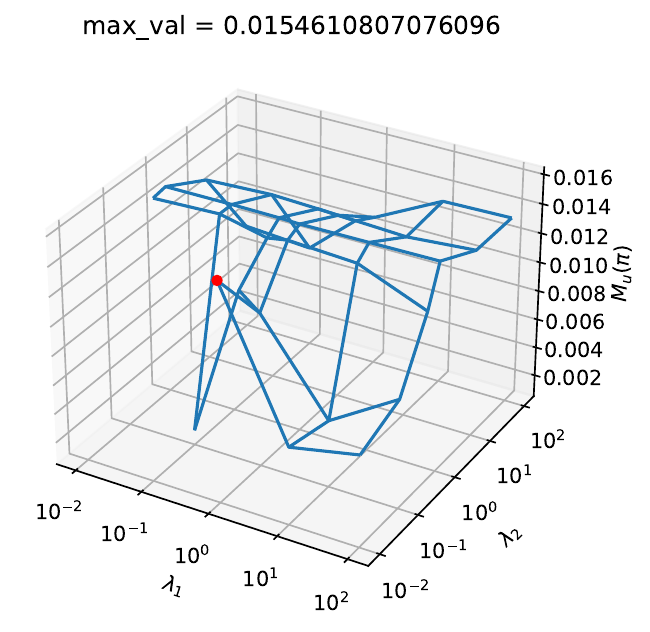}   
    \includegraphics[width=.45\textwidth]{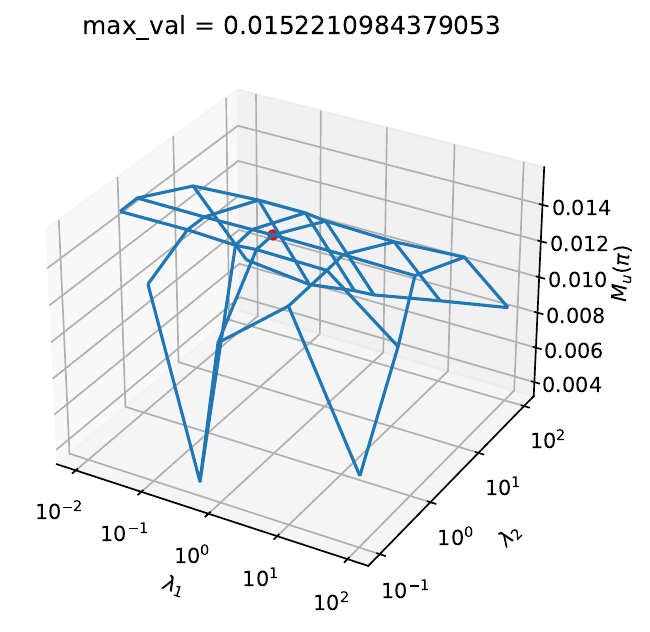}
    \includegraphics[width=.45\textwidth]{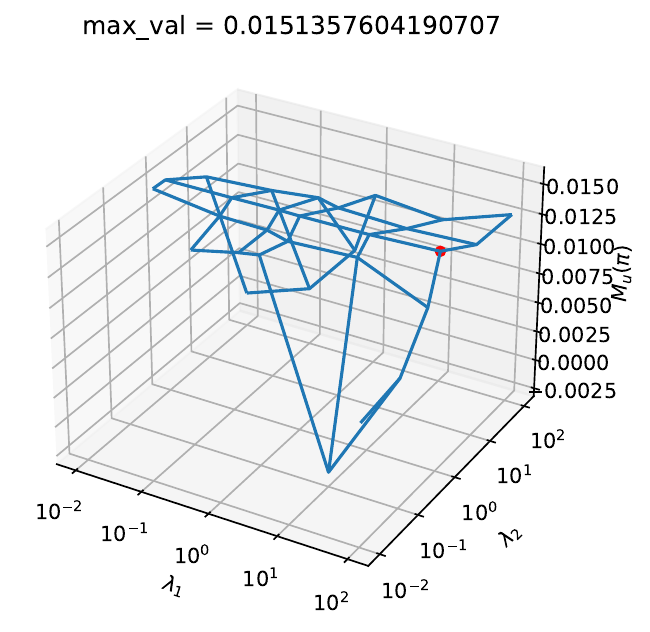}
    \caption{$M_u(\piNN_{\text{fully robust}})$ when evaluated against noisy market models (left) and against noisy GARCH models (right) for different scaling parameters $\lambda_1,\lambda_2$, for the utility functions $u_{2}$ (top), $u_1$ (middle) and $u_{1/2}$ (bottom). Results are computed on an independent test set.}
    \label{fig:M surf plot}
\end{figure}

In \Cref{fig:M surf plot} (left) we show the results of a grid search over the scaling factors  $\lambda_1,\lambda_2\in\{0.01, 0.1, 0.5, 1, 10, 100\}$, which scale the penalties $R_1$ and $R_2$ for the quadratic covariation $l_{\text{QCV}}$ and for the average relative return $l_{\text{ARR}}$, respectively, for the 3 utility functions $u_{1/2}, u_1, u_{2}$.
Moreover, in \Cref{tab:results realistic settings} we see that for all 3 utility functions, the best combination of $\lambda_1,\lambda_2$ leads to the pooled minimum expected utilities $M_{u}(\piNN_{\text{fully robust}})$, which are better (higher) than the baseline of investing everything in the risk-free bank account (given by $u(X_0 \, e^{rT})$) and higher than the results of strategies trained in the non-robust setting, corresponding to $\lambda_1 = \lambda_2 = \infty$, as well as in the half-robust setting, where only the volatility is robustified, corresponding to $\lambda_2=\infty$ (here the maximum is always attained at $\lambda_1=0.01$). 
\begin{table}[!t]
\caption{Numerical results of reference values and trained strategies evaluated against noisy market models for the different utility functions. In all cases the fully robust model with best scaling parameters outperforms all other strategies.}
\label{tab:results realistic settings}
\begin{center}
\begin{tabular}{l | r r r r}
\toprule
   & $\displaystyle \max_{\lambda_1, \lambda_2} M_u(\piNN_{\text{fully rob}})$   & $\displaystyle \max_{\lambda_1} M_u(\piNN_{\text{vola rob}})$ & $\displaystyle M_u(\piNN_{\text{non rob}})$   & $u(X_0 \, e^{rT})$  \\ \midrule 
$u_{1/2}$     &  0.01522 & 0.01465 & 0.00771 & 0.01506\\
$u_{1}$     &  0.01509 & 0.01464 & 0.00901 & 0.01500\\
$u_{2}$     &  0.01492 & 0.01391 & 0.01261 & 0.01488\\
\bottomrule
\end{tabular}
\end{center}
\end{table}
In particular, we see that the robust training leads to reasonable strategies, when using the appropriate penalty scaling parameters $\lambda_1$ and $\lambda_2$ for the given noisy evaluation market models and utility function. 
We also see that for all utility functions the optimal scaling parameters are close ($(\lambda_1,\lambda_2)=(0.5, 10)$ for $u_{2}$, $(\lambda_1,\lambda_2)=(0.5,0.5)$ for $u_1$, $(\lambda_1,\lambda_2)=(1, 1)$ for $u_{1/2}$).
In real world settings this means that one has to first decide or understand against which type or magnitude of uncertainty one wants to robustify and then has to find appropriate penalty scaling factors for the respective setting, e.g., via grid search.
Note that when comparing (minimal) expected utilities across different utility functions, it is important to keep in mind that larger utility values do not imply favorable distributions of the terminal wealth $X_T$, since different utility functions assign different values to the same terminal wealth (cf.\ \Cref{fig:utility functions}).

\begin{figure}[!htb]
    \centering
    \includegraphics[width=.49\textwidth]{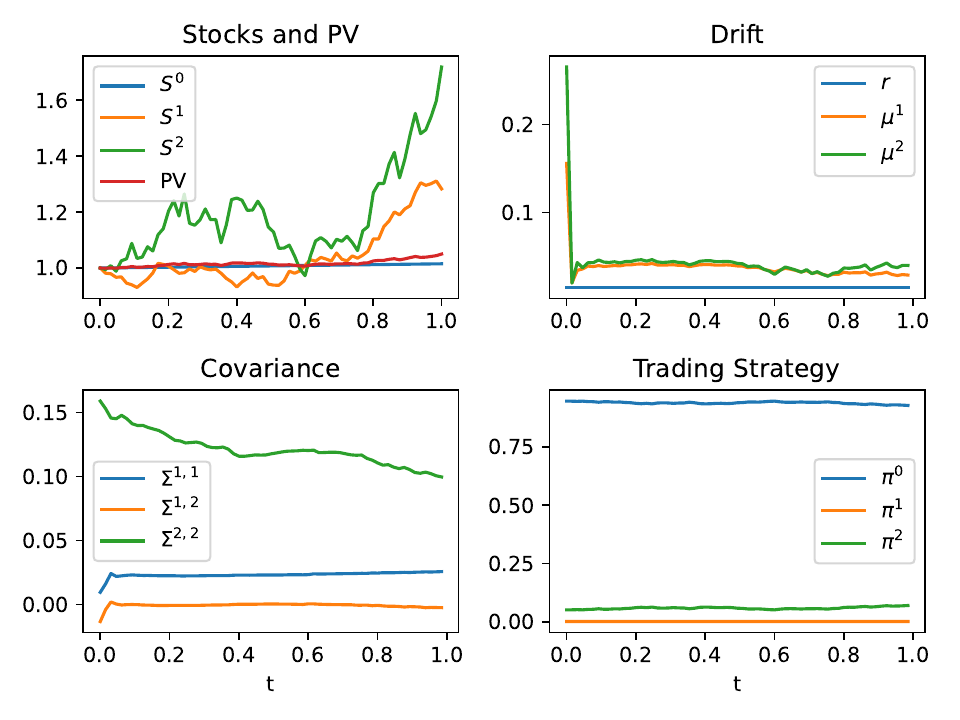}
    \includegraphics[width=.49\textwidth]{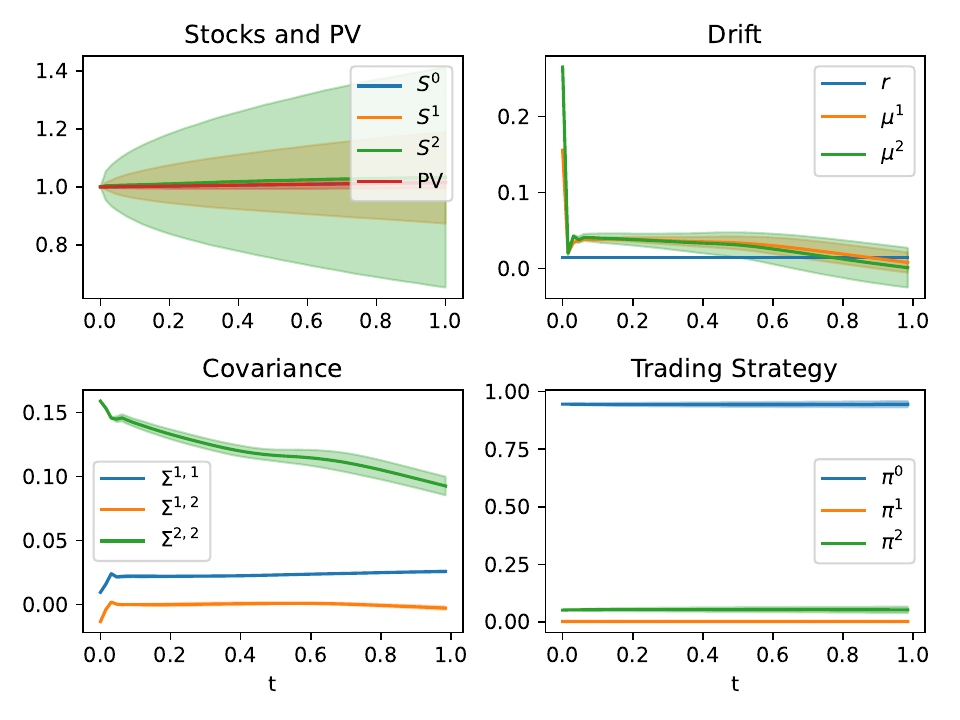}
    \includegraphics[width=.49\textwidth]{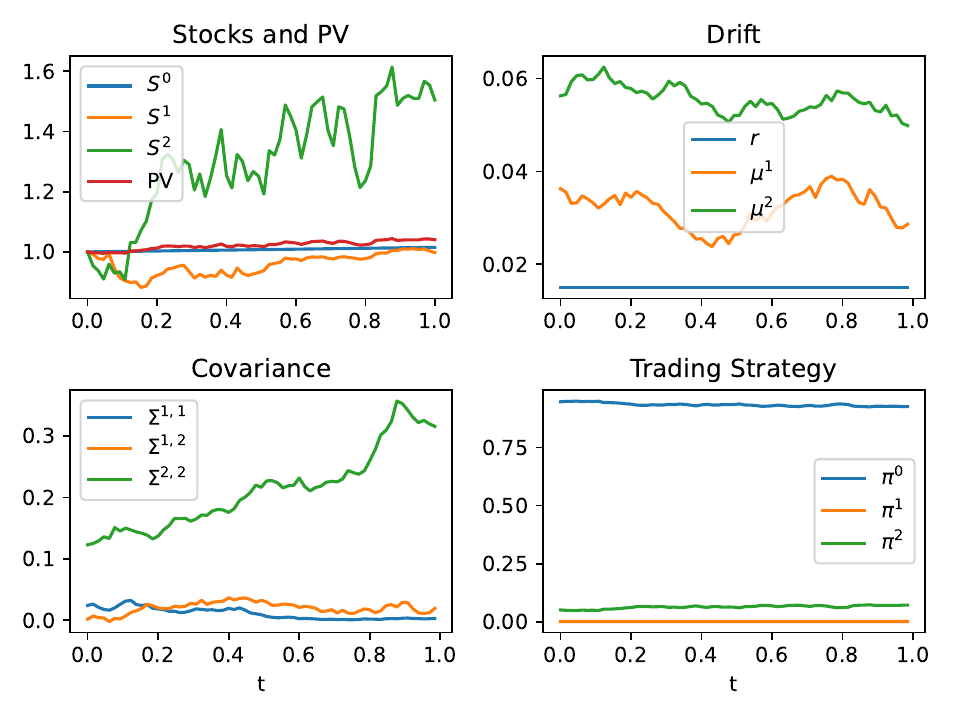}
    \includegraphics[width=.49\textwidth]{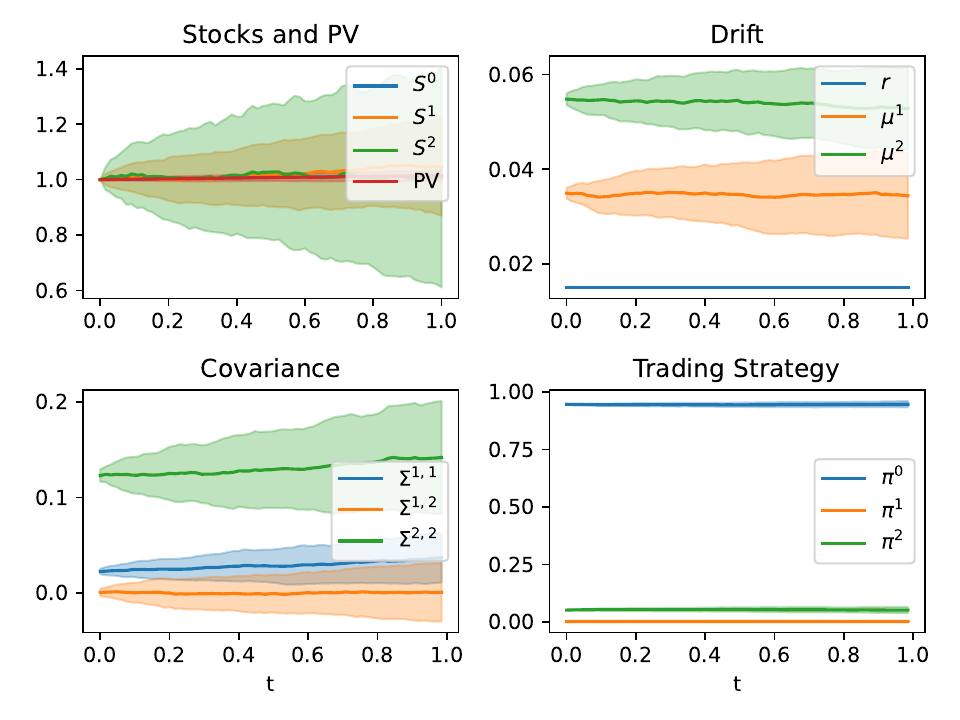}
    \caption{One test path samples (left) and mean $\pm$ std over all path samples (right) when evaluating the trained strategy for $u_{1/2}$ utility against the discriminator market model, i.e. the trained worst case market, (top) and against one noisy market model (bottom). In each of the plots we see the evolution of the stocks and portfolio value, the drift and covariance coordinates and the trading strategy's coordinates, which all influence each other in the case of the discriminator market model.}
    \label{fig:path and average plots power0.5}
\end{figure}

In \Cref{fig:path and average plots power0.5}, we show test sample paths and mean $\pm$ standard deviations over all sample paths for the best strategy with utility $u_{1/2}$ evaluated against the trained market discriminator model and against one cumulative noisy market scenario.  
Especially in the average plots (the right panels of \Cref{fig:path and average plots power0.5}), we see that the trained market and the noisy market behave differently.
Nevertheless, the trained strategy works well when evaluated against the noisy market model. In particular, the average behaviour of the strategy and the resulting average portfolio value evolution are very similar in both cases, even though the market behaves differently.

Similarly to before in \Cref{sec:Setting with explicit Solution,sec:MarketWithTransactionCosts},  FFNNs performed on par with RNNs (tested with the utility function $u_1$). This finding suggests that also in this more realistic and complex setup the optimal strategy is \emph{not} path-dependent.

\paragraph{Evaluation with Noisy GARCH models.}
Since evaluating the trained strategies against noisy GARCH models can be quite different from evaluating them against noisy market models around the reference market, we retrain the same models using noisy GARCH models for model selection. In particular, the entire training (including the discriminator and its penalties) is identical, except for the validation during training which is now done with \eqref{equ:early stopping metric} and noisy GARCH models.
For sampling the noisy GARCH models for model evaluation, we use the following standard deviations.
For the Gaussians around the fitted parameters, we use $0.75$ times their standard errors, and we set the standard deviation for the correlations to $0.0075$. 
As previously in the evaluation with noisy market models, we use larger standard deviations to sample the noisy GARCH models for model selection (i.e., early stopping during training) to counteract the too optimistic view of the validation metric \eqref{equ:early stopping metric}. In particular, we use a factor of $1$ on the standard errors of the fitted parameters, and set the standard deviation for the correlations to $0.01$.
For early stopping we use \eqref{equ:early stopping metric} with $B_{\text{val}}=500$ and for validation we use \eqref{equ:eval metric} with $B_{\text{test}}=5000$ and $n_{\text{pool}}=1000$ (as before).
\begin{table}[!t]
\caption{Numerical results of reference values and trained strategies evaluated against noisy GARCH models for the different utility functions. In all cases the fully robust model with best scaling parameters outperforms all other strategies.}
\label{tab:results realistic settings GARCH}
\begin{center}
\begin{tabular}{l | r r r r}
\toprule
   & $\displaystyle \max_{\lambda_1, \lambda_2} M_u(\piNN_{\text{fully rob}})$   & $\displaystyle \max_{\lambda_1} M_u(\piNN_{\text{vola rob}})$ & $\displaystyle M_u(\piNN_{\text{non rob}})$   & $u(X_0 \, e^{rT})$  \\ \midrule 
$u_{1/2}$   &  0.01514 & 0.01407 & 0.01372 & 0.01506\\
$u_{1}$     &  0.01546 & 0.01423 & 0.01388 & 0.01500\\
$u_{2}$     &  0.01496 & 0.01393 & 0.01413 & 0.01488\\
\bottomrule
\end{tabular}
\end{center}
\end{table}
In \Cref{fig:M surf plot} (right) we show the results of a grid search over scaling factors $\lambda_1, \lambda_2 \in \{ 0.01, 0.5, 0.1, 1., 10, 100 \}$ for the path-dependent penalties w.r.t.\ the quadratic covariation $l_{\text{QCV}}$ and the average relative return $l_{\text{ARR}}$, for the utility functions $u_{1/2}, u_1,$ and $ u_2$.
In \Cref{tab:results realistic settings GARCH}, we see that for all 3 utility functions, the best combination of $\lambda_1,\lambda_2$ leads to the pooled-minimum expected utilities $M_{u}(\piNN_{\text{fully robust}})$ that exceed the baseline of investing everything in the risk-free bank account (given by $u(X_0 \, e^{rT})$). Moreover, $\displaystyle \max_{\lambda_1, \lambda_2}M_{u}(\piNN_{\text{fully robust}})$ is also higher than the results of strategies trained in the non-robust setting, corresponding to $\lambda_1 = \lambda_2 = \infty$, as well as in the half-robust setting, where only the volatility is robustified, corresponding to $\lambda_2=\infty$ (which is always attained for $\lambda_1=0.01$). 

In total, the evaluation against the noisy GARCH models shows that the robust model outperforms the others, when choosing the scaling factors correctly.
Interestingly, the fully robust model for $u_1$ leads to a slightly higher value of $M_{u}(\piNN_{\text{fully robust}})$ than for $u_{1/2}$. This is surprising, since $u_{1/2}$ dominates $u_1$ and one would expect this monotonicity to translate to the metric.\footnote{In particular, using the strategy that results from the optimization w.r.t.\ $u_1$ in the evaluation with $u_{1/2}$ has to lead to the same or higher expected utility than the evaluation w.r.t.\ $u_1$.} As we can see in \Cref{fig:M surf plot}, the best combination of scaling factors is always attained at a boundary point of the grid. Hence, it is expected that better scaling factors could be found, when further enlarging the grid. In line with this, we expect that then also the optimal combination for $u_{1/2}$ leads to a larger value of $M_{u}(\piNN_{\text{fully robust}})$ than for $u_1$.

\paragraph{Evaluation on Reference Market Model.}
\begin{figure}[!htb]
    \centering
    \includegraphics[width=.49\textwidth]{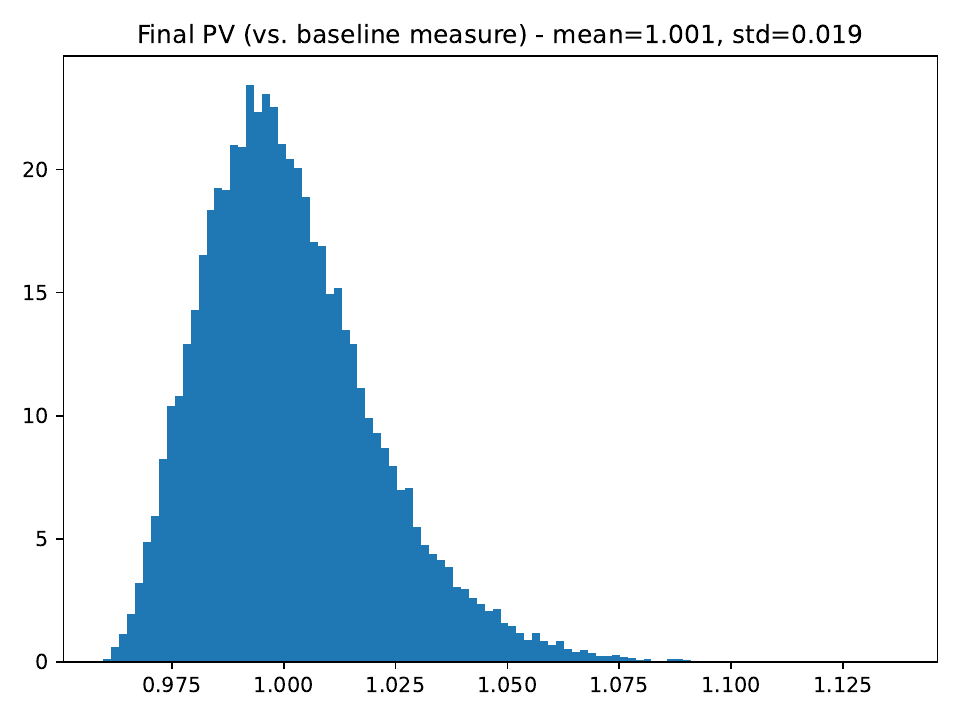}
    \includegraphics[width=.49\textwidth]{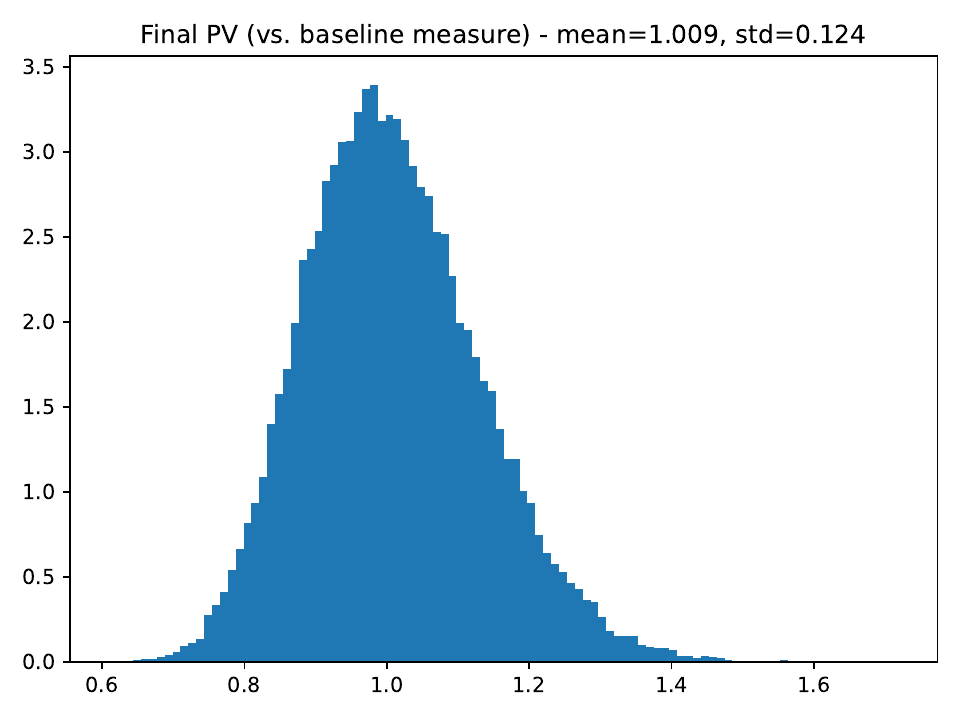}
    \includegraphics[width=.49\textwidth]{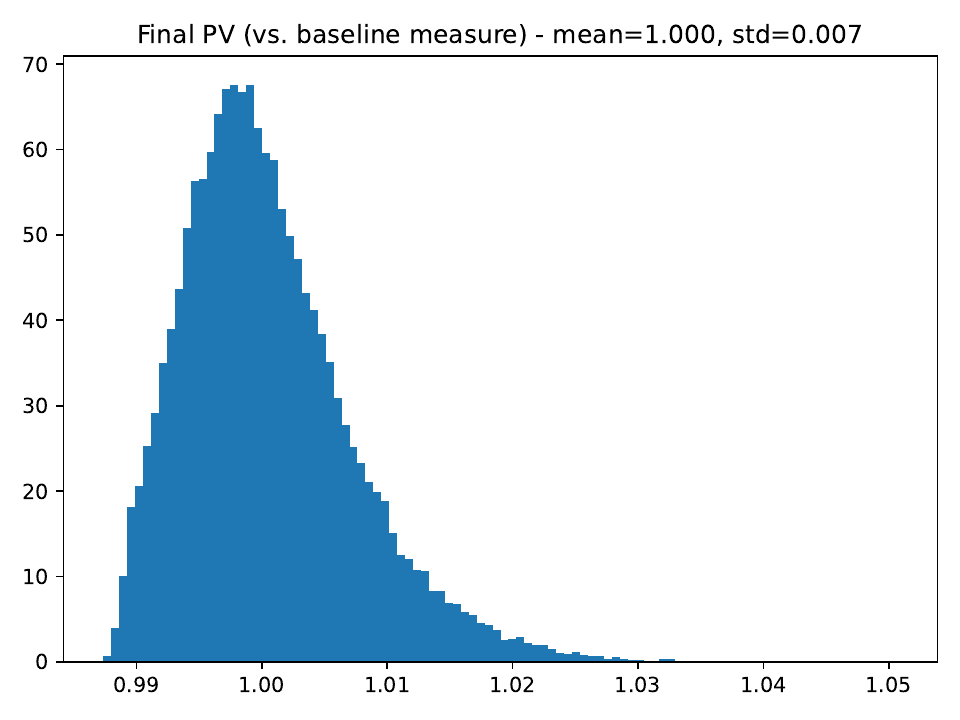}
    \includegraphics[width=.49\textwidth]{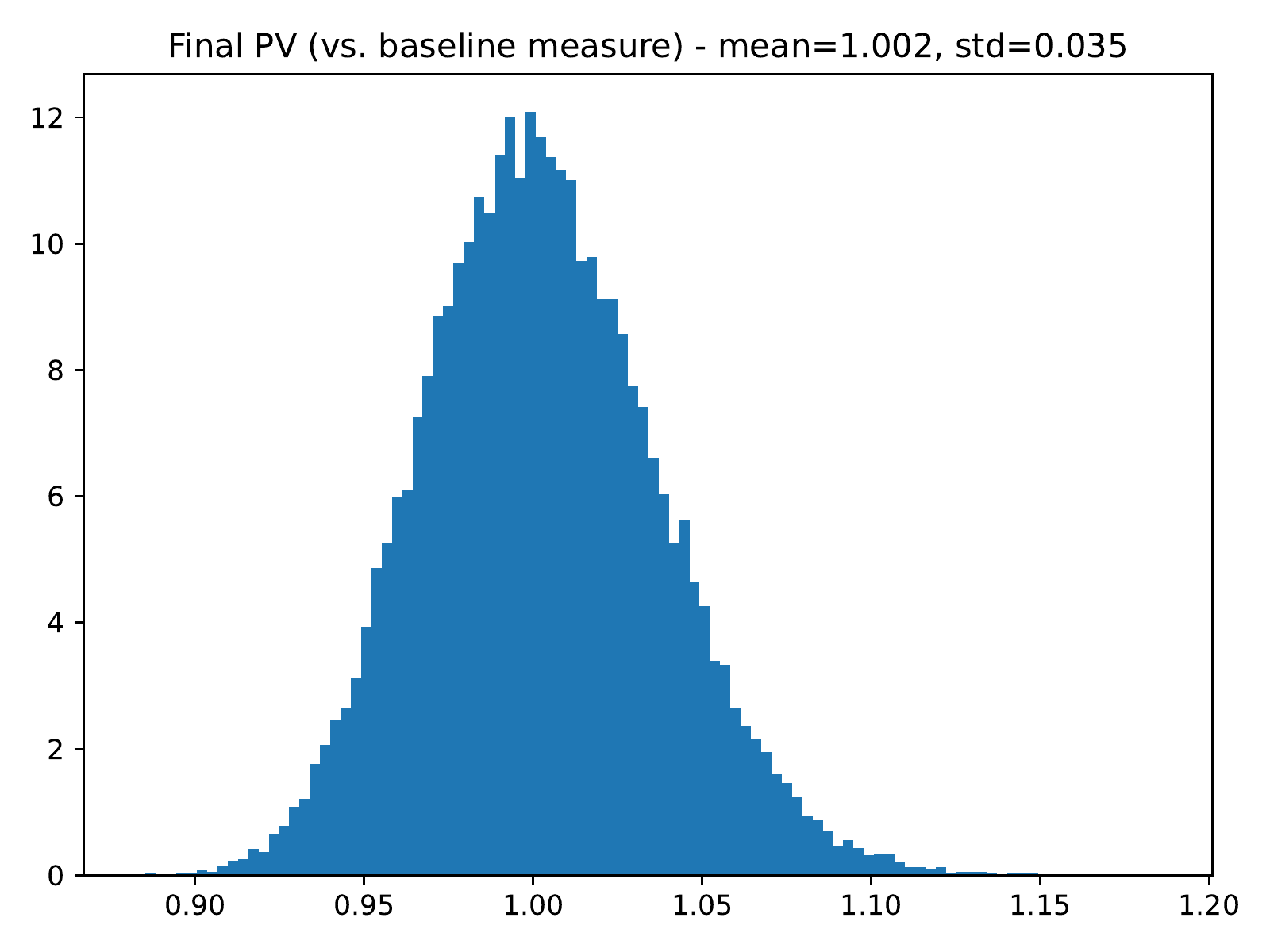}
    \caption{Histograms of discounted terminal portfolio value when evaluating a trained strategy against a constant baseline measure $(\tilde\mu,\tilde\Sigma)$ for fully robust (left) and non-robust (right) models, with utility function $u_{1/2}$  (top) and $u_2$ (bottom). 
    Expected Utilities (left to right, top to bottom): 0.0164, 0.0203, 0.0153, 0.0160.
    }
    \label{fig:histogramsTerminalPV}
\end{figure}
In \Cref{fig:histogramsTerminalPV} we compare the distribution of discounted terminal wealth $e^{-rT} \, X_T$ resulting from trading with the fully robust and non-robust strategies of \Cref{tab:results realistic settings} $\piNN_{\text{fully rob}}$ and $\piNN_{\text{non rob}}$, respectively, obtained by (robust) portfolio optimization with the utility functions $u_2$ and $u_{1/2}$. 
We see that the trading with robust strategies leads to lower mean terminal portfolio value (PV), however, also manages to reduce the termianal PV's standard deviation. Hence, even though there is some cost for the robustification, it also reduces the risk of the strategy. 
Additionally, the robust strategies have the advantage of being right-skewed (large profits are more likely to happen than large losses) compared to the non-robust ones which are not skewed (large profits and large losses are similarly likely to happen).
Except for the skew, we see that changing from the non-robust (top-right) to the robust (top-left) strategy has a qualitatively similar effect as changing from a more risk-seeking (top-right) to a more risk-avers (bottom-right) utility function. Doing both (from top-right to bottom-left) increases the effect.

\subsection{(Non-)Robust Utility Optimization with Small Trading Cost}\label{sec:Robust Utility Optimization with Small Trading Cost}
In this section, we study in more detail the effect of small, proportional trading costs on optimal investments in both robust and non-robust utility optimization.
In this paper, we so far tested how a GAN-based algorithm (cf. \Cref{sec:DeepRobustUtilityOptimization}) can be leveraged to recover known explicit solutions to robust utility optimization (\Cref{sec:Setting with explicit Solution}) and enable us to go beyond simple, friction-less settings to find robust strategies in presence of proportional trading costs and for general power utilities (\Cref{sec:MarketWithTransactionCosts,sec:RealisticSettingForStockMarket}).
While we saw benefits of GAN-based trading strategies in expected utilities under noisy market evaluations in \Cref{sec:MarketWithTransactionCosts,sec:RealisticSettingForStockMarket}, in this \Cref{sec:Robust Utility Optimization with Small Trading Cost}, we investigate potential benefits of the learned trading strategies compared to other well known, asymptotically optimal investments under vanishing trading costs.

In non-robust portfolio optimization, such an analysis of optimal trading strategies in presence of (vanishingly) small trading costs has been performed in \citet{muhlekarbe2017primer}. For general power utilities of the form $\tilde u_p(x) = \tfrac{x^{1-p}}{1-p}$, for $p \in (0,1)$ (note that this definition slightly differs from our previous definition in \Cref{eq:ourpowerutility}), \cite{muhlekarbe2017primer} derive closed form approximations of asymptotically optimal trading strategies in presence of small trading costs for Black--Scholes and more general mean-reverting financial markets. 
In this section, we take up their discussion by comparing our GAN-based robust and non-robust strategies to the asymptotically optimal strategy derived in \citet{muhlekarbe2017primer} for Black--Scholes markets in \Cref{sec:Comparison of the Reference and the GAN Strategy}.
Then, in \Cref{subsubsec:Comparison in Markets with Heavy tails}, we compare those strategies in one-dimensional markets consisting of a risky asset with heavy tailed log-returns.

\subsubsection{Comparison in Black--Scholes Markets}\label{sec:Comparison of the Reference and the GAN Strategy}

In this section, we expand on the analysis of optimal investments under vanishing transaction costs  of \cite{muhlekarbe2017primer} in two ways. First, for a Black--Scholes market with small trading costs and power utility, we compare the optimal strategy derived in \cite[Section 5.1]{muhlekarbe2017primer} to the non-robust GAN-based strategy trained in the same financial environment.
Second, we compare those strategies to fully robust GAN-based strategies in noisy market environments to investigate additional effects of robustification in the more realistic setting of market uncertainty. 

In all subsequent analyses of this section, we consider a one-dimensional Black--Scholes market with risk-free asset $B_t = e^{rt}$ and risky asset $S$ given by \eqref{eq:assetdynamics-experiments} with constant parameters $\mu_t = \tilde\mu \in\R, \sigma_t = \tilde\sigma \in\R_{\ge0}$.
To match the definition in \cite[Section 2.1]{muhlekarbe2017primer}, we define $\tilde\mu = \tilde\mu_S + r$, where $\tilde\mu_S \in \R$ is the expected excess return (of the risky asset over the risk-free one). Moreover, we assume to have only proportional trading costs $c_{\text{prop}} > 0$ (i.e., we set $c_{\text{base}}=0$) and a power utility function $\tilde u_p$ for $p \in (0,1)$.
Then \cite[Section~4.5 and~5.1]{muhlekarbe2017primer} yield that the optimal (proportional) trading strategy in the case without transaction costs is given by 
\[ \pi_{\text{ntc}}^{\text{ref}} = \tfrac{\tilde\mu_S}{p \tilde\sigma^2}. \] 
Let $Y_t := \pi_t \, X_t$ be the wealth at time $t$ that is invested in the risky asset and $H_t := \pi_t \tfrac{X_t}{S_t}$ be the number of risky assets currently held. 
Then the asymptotically optimal trading strategy with trading costs is described by the asymptotic no-trade region 
\begin{equation}\label{equ:no trade region}
    \operatorname{NT} = \left\{ \pi \in \R \, \Big\vert \, \vert \pi - \pi_{\text{ntc}}^{\text{ref}} \vert \leq  c_{\text{prop}}^{1/3} \Delta\pi  \right\},  
\end{equation}
with
\begin{equation*}
    \Delta\pi = \left( \frac{3}{2p} \left( \pi_{\text{ntc}}^{\text{ref}} (1-\pi_{\text{ntc}}^{\text{ref}}) \right)^2 \right)^{1/3}.
\end{equation*}
In particular, whenever the current proportion $Y_t/X_t = \pi_t$ invested in the risky asset lies in the no-trade region $\operatorname{NT}$, the asymptotically optimal strategy with trading costs $\pi^{\text{ref}}$ tells not to trade, i.e., to keep the number of risky assets $H_t$ constant. On the other hand, whenever $Y_t/X_t = \pi_t$ is outside the no-trade region, the asymptotically optimal strategy is to minimally adjust $H_t$ such that $\pi^{\text{ref}}_t$ lies in the no-trade region again. Therefore, we have
\begin{equation*}
    \pi^{\text{ref}}_t = \begin{cases}
         \tfrac{H_{t} S_t}{X_t},  & \text{if } \tfrac{Y_t}{X_t} \in \operatorname{NT}, \\
        \pi_{\text{ntc}}^{\text{ref}} + c_{\text{prop}}^{1/3} \Delta\pi, & \text{if } \tfrac{Y_t}{X_t} > \max( \operatorname{NT} ), \\
        \pi_{\text{ntc}}^{\text{ref}} - c_{\text{prop}}^{1/3} \Delta\pi, & \text{if } \tfrac{Y_t}{X_t} < \min(\operatorname{NT} ),
    \end{cases}
\end{equation*}
and $dH_t=0$, if ${Y_t}/{X_t} \in \operatorname{NT}$.
\paragraph{Experiment setup}
In our experiments we use a similar setup as before with $p=0.5$, $c_{\text{prop}} = 1\%$, $X_0 =1$, $S_0=1$, $\tilde\sigma=35\%$, $\delta_t = 1/65$ with $65$ steps, i.e., $T=1$, and two different regimes for the drift and risk-free rate $(\tilde\mu_S, r) \in \{ (4\%, 1.5\%), (7\%, 3\%)\}$, hence, $\tilde\mu \in \{ 5.5\%, 10\% \}$.
For the robust training we use the path-dependent penalties of \Cref{sec:Path-dependent Penalties} and grid search over $\lambda_1, \lambda_2 \in \{ 0.01, 0.5, 0.1, 1., 10, 100 \}$. For validation of robust models, we consider noisy market models around the reference market, with the same setup as in the corresponding paragraph in \Cref{sec:Experimental Results}. For the non-robust model, validation is done without noise, i.e., with the reference market with characteristics $\tilde\mu$ and $\tilde\sigma$ by approximating \eqref{eq:explicitSigExpectedUtility} with $B_{\text{val}} = 40K$ independent validation sample paths. For each experiment, we run a range of models with various architectures and select the model that performed best in the validation for final evaluation on a separate test set. In the evaluation, we approximate the metric \eqref{equ:eval metric} with $n_{\text{pool}}=1000$ and $B_{\text{test}}=40K$, and the metric \eqref{eq:explicitSigExpectedUtility} with $B_{\text{test}}=100K$.
Again, we did not see an advantage of using RNNs compared to standard FFNNs, therefore all results are reported for FFNNs.

\paragraph{Results}
In \Cref{tab:results small tans costs}, we show the comparison of the reference strategy $\pi^{\text{ref}}$ and the trained strategies $\piNN$ evaluated against the noisy market models and against the reference markets (with highest, i.e., best values per column in bold). For the fully robust strategy $\piNN_{\text{fully rob}}$, we show the results of the best combination of scaling factors, given by $\lambda_1=10, \lambda_2=1$ (for $\tilde\mu=5.5\%$ and $r=1.5\%$), and by $\lambda_1=10, \lambda_2=10$ (for $\tilde\mu=10\%$ and $r=3\%$), respectively.
\begin{table}[!tb]
\caption{
Numerical results of reference strategy and trained GAN strategies evaluated against noisy market models with \eqref{equ:eval metric} and against the reference market with \eqref{eq:explicitSigExpectedUtility}.
Additionally, the discounted value at risk $\operatorname{VaR}_\alpha$ for $\alpha = 5\%$ is computed in the noise-free reference markets, corresponding to the quantiles of the terminal portfolio value plotted in \Cref{fig:terminal PV ref strategy small trading costs}.
}
\label{tab:results small tans costs}
\begin{center}
\begin{tabular}{l | r r r | r r r}
\toprule
\multirow{2}{*}{}&\multicolumn{3}{c|}{$\tilde\mu=5.5\%$, $r=1.5\%$}&\multicolumn{3}{c}{$\tilde\mu=10\%$, $r=3\%$}\\
\cmidrule{2-7}
   & $\displaystyle M_{\tilde u_p}(\pi)$   & $E_{\tilde u_p}(\pi, \tilde\mu, \tilde\sigma)$ & $\operatorname{VaR}_\alpha$ & $\displaystyle M_{\tilde u_p}(\pi)$& $E_{\tilde u_p}(\pi, \tilde\mu, \tilde\sigma)$  & $\operatorname{VaR}_\alpha$ \\ \midrule[\heavyrulewidth]
   $\pi^{\text{cash}}$
&\multicolumn{2}{c}{$2.0151$} & 0 &\multicolumn{2}{c}{$2.0302$} & 0 \\
$\pi^{\text{ref}}$   &  2.0080 & {\bf 2.0221} & 0.28 & 2.0231& {\bf 2.0602} & 0.47 \\
$\piNN_{\text{non rob}}$     &  2.0117 & {\bf 2.0221} & 0.22 & 2.0279 & {\bf 2.0605} & 0.44\\
$\piNN_{\text{fully rob}}$     &  {\bf 2.0154 }& 2.0160& 0.02& {\bf 2.0406} & 2.0541 & 0.25 \\
\bottomrule
\end{tabular}
\end{center}
\end{table}
\begin{figure}[!b]
    \centering
    \includegraphics[width=.49\textwidth]{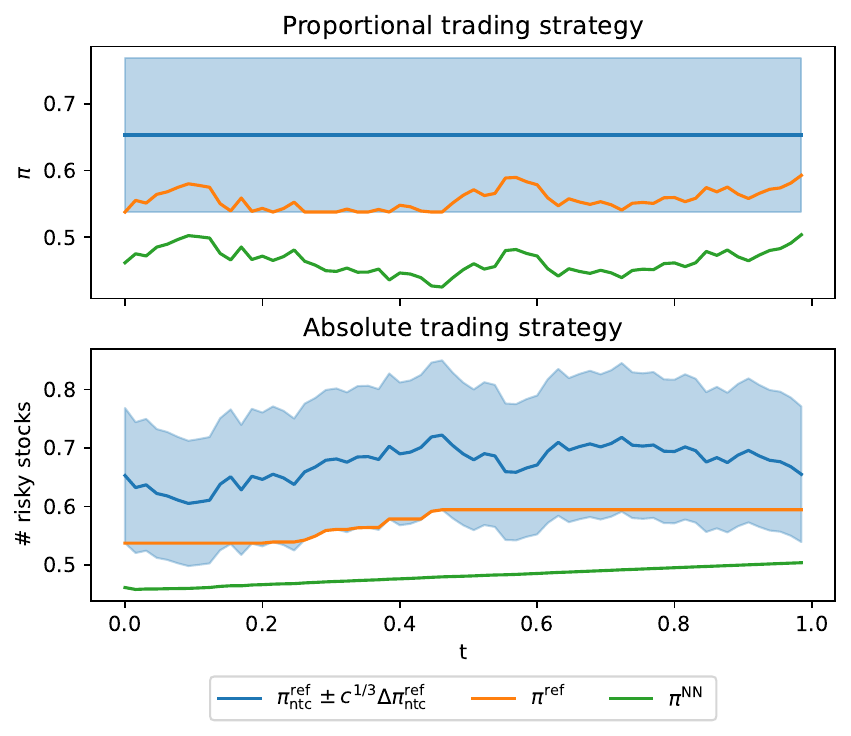}
    \includegraphics[width=.49\textwidth]{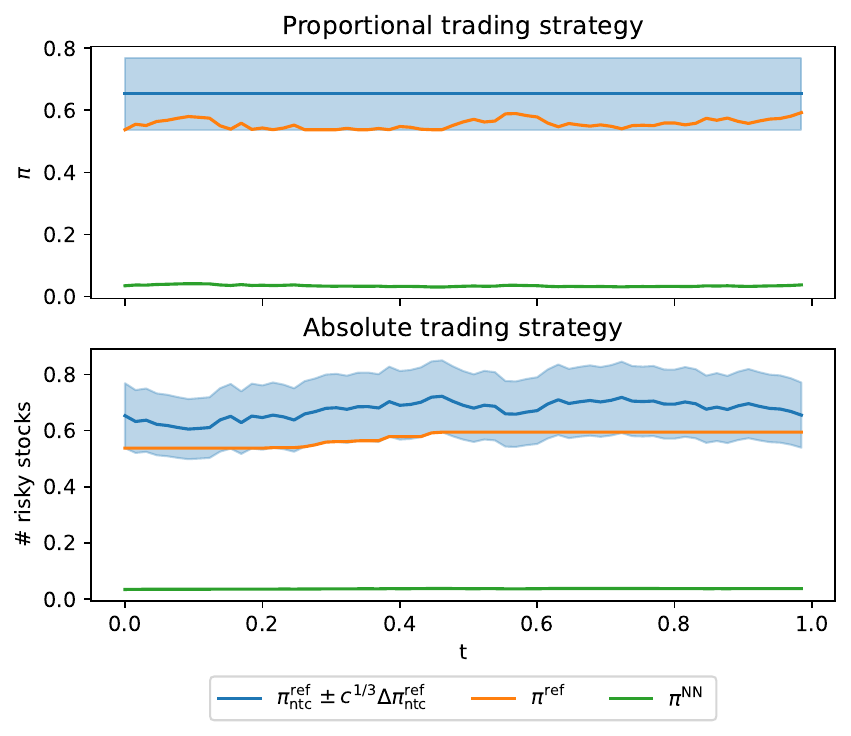}
    \includegraphics[width=.49\textwidth]{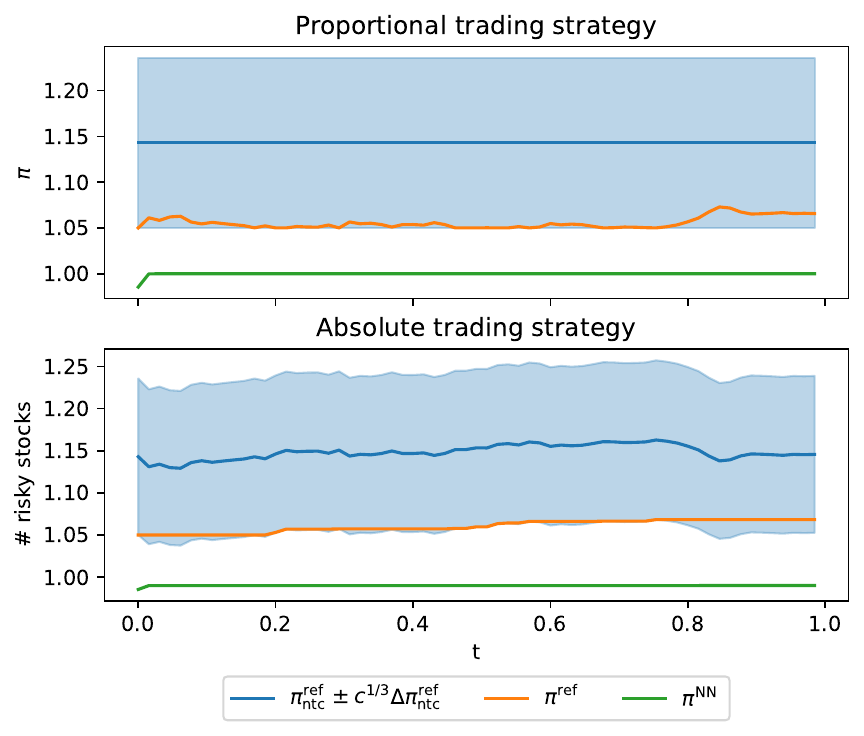}
    \includegraphics[width=.49\textwidth]{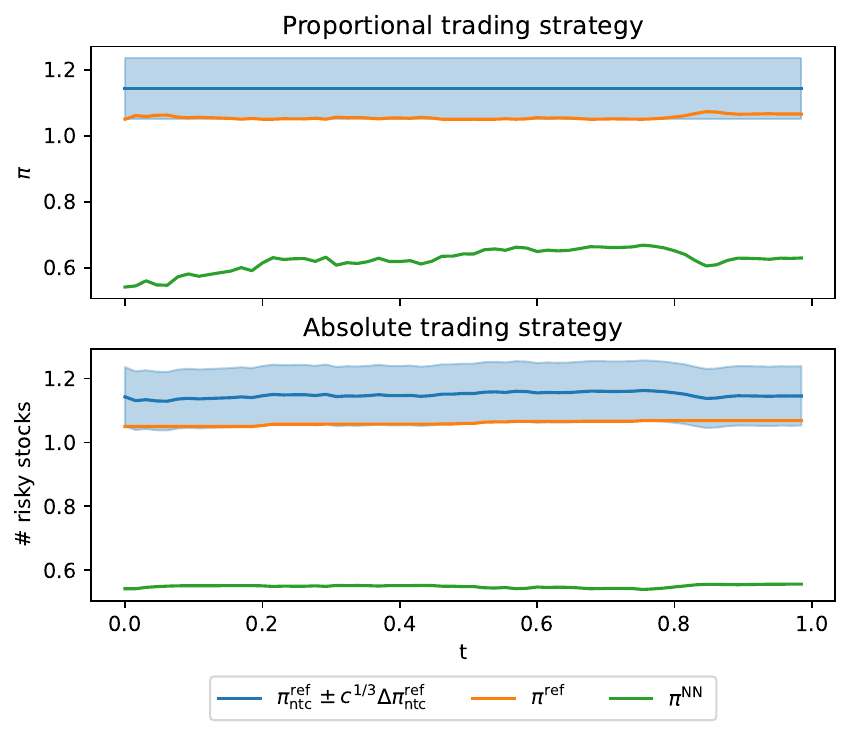}
    \caption{
    Test paths of the optimal reference strategy without trading costs $\pi_{\text{ntc}}^{\text{ref}}$ together with the no-trade-region, the asymptotically optimal reference strategy with trading costs $\pi^{\text{ref}}$ and of the non-robustly trained GAN strategy $\piNN_{\text{non rob}}$ (left) and robustly trained GAN strategy $\piNN_{\text{fully rob}}$ (right). The top row shows paths in the market with $\tilde\mu=5.5\%$ and $r=1.5\%$, and the bottom row with $\tilde\mu=10\%$ and $r=3\%$.
    In each of the four sub-figures, the proportional ($\pi_t$; top) and the absolute ($H_t$; bottom) strategies are plotted.
    For the reference strategy we see that $H_t^{\text{ref}}$ stays constant within the no-trade region, which implies a changing portfolio weight $\pi^{\text{ref}}_t$, due to the change in price $S_t$ and portfolio value $X_t$. 
    }
    \label{fig:ref strategy small trading costs}
\end{figure}
\begin{figure}[!b]
    \centering
    \includegraphics[width=.49\textwidth]{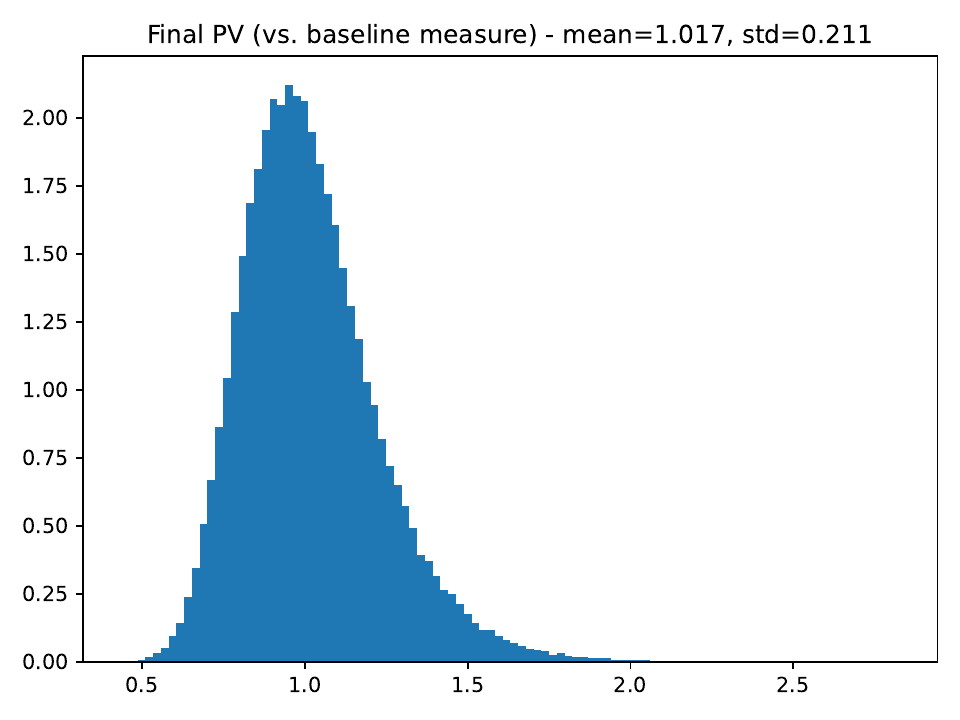}
    \includegraphics[width=.49\textwidth]{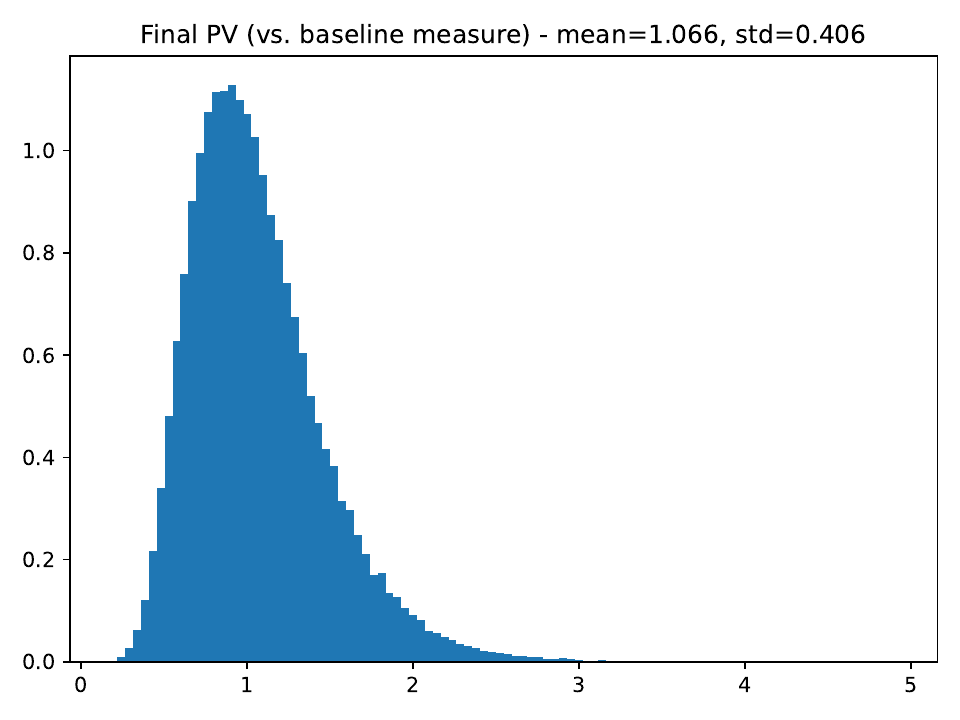}
    \includegraphics[width=.49\textwidth]{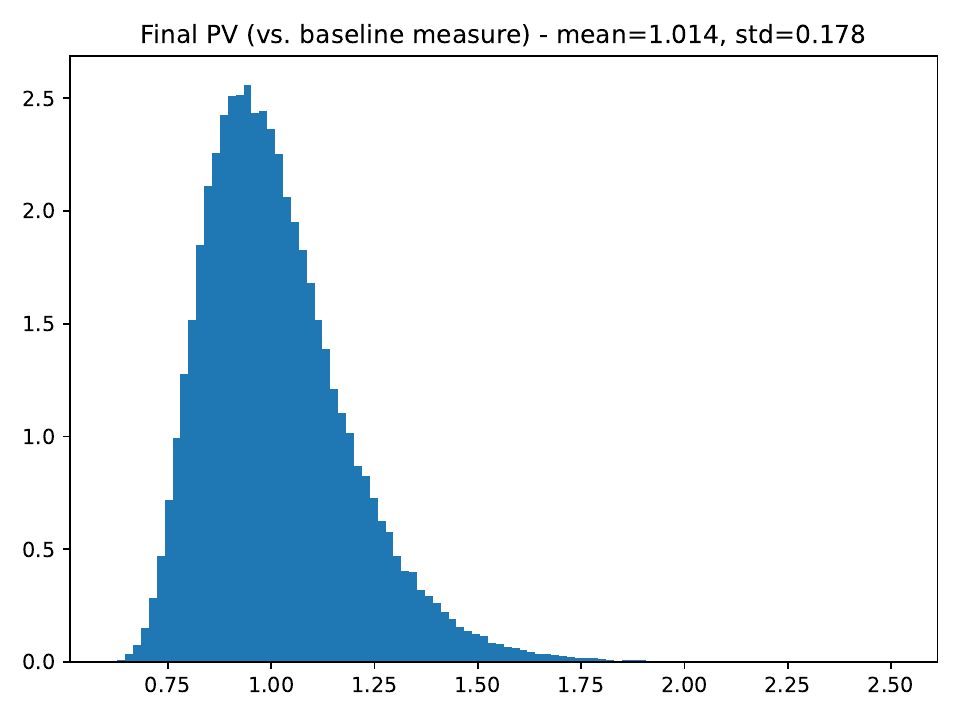}
    \includegraphics[width=.49\textwidth]{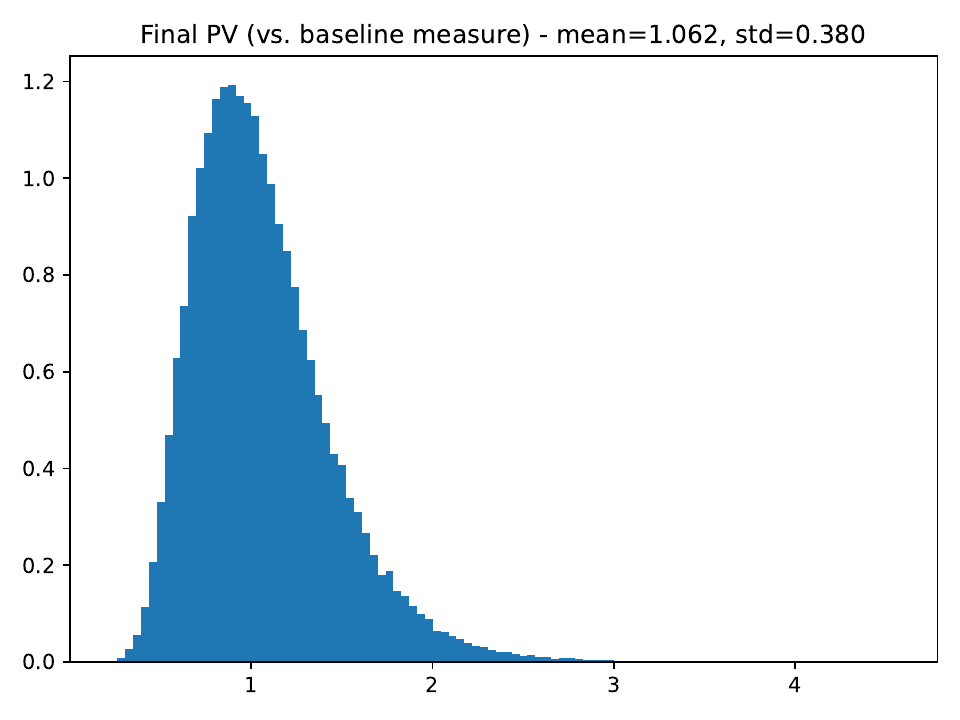}
    \includegraphics[width=.49\textwidth]{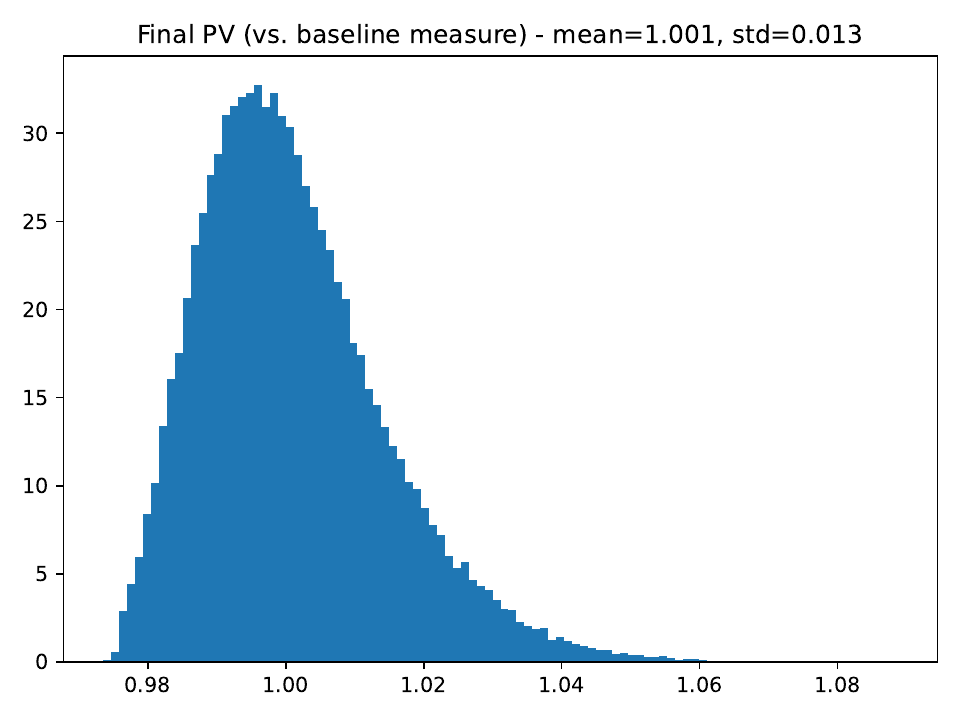}
    \includegraphics[width=.49\textwidth]{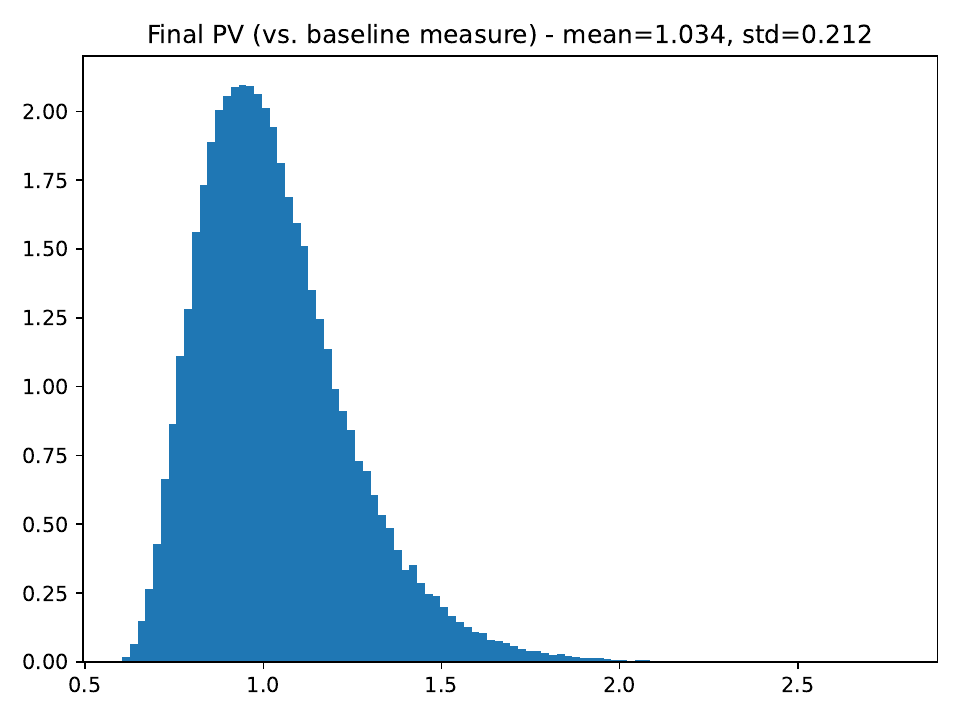}
    \caption{
    Histograms of discounted terminal portfolio values $X_T$  for the asymptotically optimal reference strategy with trading costs $\pi^{\text{ref}}$ (top), the non-robustly trained GAN strategy $\piNN_{\text{non rob}}$ (middle) and the robustly trained GAN strategy $\piNN_{\text{fully rob}}$ (bottom) in the noise-free market with $\tilde\mu=5.5\%$ and $r=1.5\%$ (left), and with $\tilde\mu=10\%$ and $r=3\%$ (right).
    }
    \label{fig:terminal PV ref strategy small trading costs}
\end{figure}
In the non-robust case, we see that the GAN strategy $\piNN_{\text{non rob}}$ performs on par with the reference strategy $\pi^{\text{ref}}$ in the evaluation on the reference market ($E_{\tilde u_p}$ entries of \Cref{tab:results small tans costs}). The slightly higher result for $E_{\tilde u_p}(\piNN_{\text{non rob}}, \tilde\mu, \tilde\sigma)$ in the case of $\tilde\mu=10\%, r=3\%$ is most likely an effect of time discretization and the fact that the reference strategy is only asymptotically optimal for vanishing trading cost.\footnote{The values for $E_{\tilde u_p}(\piNN_{\text{non rob}},\tilde\mu, \tilde\sigma)$ and $E_{\tilde u_p}(\pi^{\text{ref}},\tilde\mu, \tilde\sigma)$ lie within one standard deviation of the corresponding Monte Carlo error.} 

Comparing one test sample of these paths in \Cref{fig:ref strategy small trading costs}, we see that $\piNN_{\text{non rob}}$ follows a similar trajectory as $\pi^{\text{ref}}$, but it is shifted down by approximately $0.1$  in the market with $\tilde\mu=5.5\%, r=1.5\%$ (top, left sub-figure of \Cref{fig:ref strategy small trading costs}) and by approximately $0.05$ in the market with $\tilde\mu=10\%, r=3\%$ (bottom, left sub-figure of \Cref{fig:ref strategy small trading costs}). Moreover, sometimes when $\pi^{\text{ref}}$ hits the boundary of the no-trade region and subsequently stays there, $\piNN_{\text{non rob}}$ develops differently, as for instance for $t\approx(0.35,0.5)$ in the top, left sub-figure. 
Thus overall, even though we observed that the learned strategy $\piNN_{\text{non rob}}$ differs from $\pi^{\text{ref}}$, the resulting expected utilities in the reference market are nearly the same, i.e., $E_{\tilde u_p}(\piNN_{\text{non rob}}, \tilde\mu, \tilde\sigma)\approx E_{\tilde u_p}(\pi^{\text{ref}}, \tilde\mu, \tilde\sigma)$ in \Cref{tab:results small tans costs}.

However, when evaluated against the noisy market ($M_{\tilde u_p}$ entries of \Cref{tab:results small tans costs}), $\piNN_{\text{non rob}}$ outperforms the reference strategy in the second digit.  This seems reasonable, since we observed in \Cref{fig:ref strategy small trading costs} how the optimized $\piNN_{\text{non rob}}$ tells to invest about 10\% of wealth less into the risky asset than $\pi^{\text{ref}}$, and therefore acts similarly to a robustly trained strategy. Note that we observe such robustified behavior in $\piNN_{\text{non rob}}$ even though the training did not enforce it for $\piNN_{\text{non rob}}$.
The robustly trained strategy $\piNN_{\text{fully rob}}$, however, outperforms both other strategies in the evaluation with noisy markets. Moreover, $\piNN_{\text{fully rob}}$ also beats the baseline $\pi^{\text{cash}} \equiv 0$ of investing everything in the risk-free asset, whose utility is given by $\tilde u_p (X_0 \, e^{rT})$, in the noise-free and the noisy market evaluation. As we have seen before, this comes at the cost of smaller expected utility when evaluated in the reference market setting ($E_{\tilde u_p}$ entries of \Cref{tab:results small tans costs}).

This can also be seen in the histograms of the discounted terminal portfolio value $X_T$ in the noise-free reference market in \Cref{fig:terminal PV ref strategy small trading costs}. Additionally, we see there that $\piNN_{\text{non rob}}$ has slightly smaller mean than $\pi^{\text{ref}}$, while also having smaller variance and being more right skewed. For the robust strategy $\piNN_{\text{fully rob}}$ this is even more pronounced. 
In particular, $\piNN_{\text{non rob}}$ manages to achieve the same expected utility as the asymptotically optimal reference strategy $\pi^{\text{ref}}$, while substantially reducing the risk, measured for example by the discounted \emph{value at risk} $\operatorname{VaR}_\alpha$ for $\alpha=5\%$ (\Cref{tab:results small tans costs}).
Indeed, with the initial wealth $X_0=1$, $\piNN_{\text{non rob}}$ reduces the value at risk compared to $\pi^{\text{ref}}$ by $6\%$ and $3\%$ in the small and large drift regimes, respectively. Moreover, the robust strategy $\piNN_{\text{fully rob}}$ reduces the value at risk by $26\%$ and $21\%$, respectively. 

Test paths of  $\piNN_{\text{fully rob}}$ are shown in the right sub-figures of \Cref{fig:ref strategy small trading costs}. Compared to their non-robust counterparts $\piNN_{\text{non rob}}$ in the left sub-figures, the proportional investment in the risky asset $\piNN_{\text{fully rob}}$ is held at much lower levels over the entire horizon.

\begin{figure}[!b]
    \centering
    \includegraphics[width=.49\textwidth]{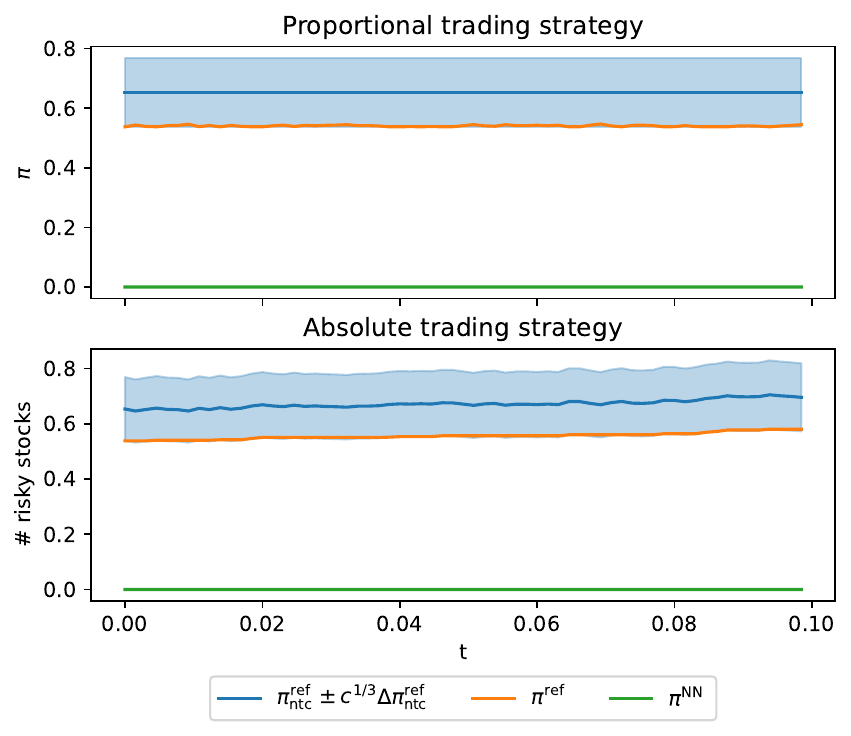}
    \includegraphics[width=.49\textwidth]{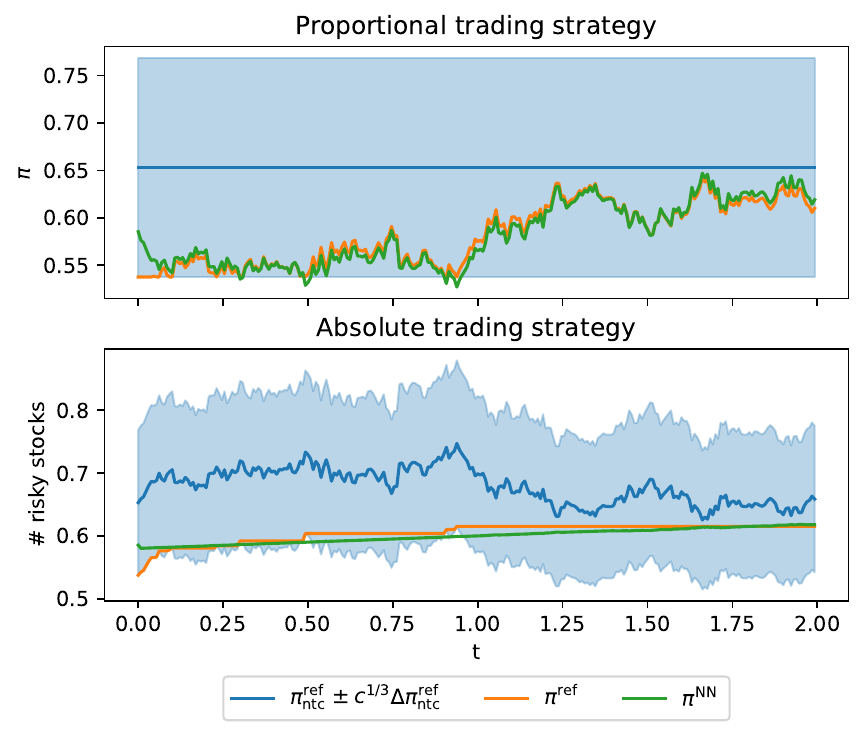}
    \caption{
    Test paths of the optimal reference strategy without trading costs $\pi_{\text{ntc}}^{\text{ref}}$ together with the no-trade-region, the asymptotically optimal reference strategy with trading costs $\pi^{\text{ref}}$ and of the non-robustly trained GAN strategy $\piNN_{\text{non rob}}$ for $T=0.1$ (left) and $T=2$ (right) in the market with $\tilde \mu =5.5\%$ and $r=1.5\%$.
    For small $T$, $\piNN_{\text{non rob}}$ learns that no investment in the risky stock is beneficial, while for large $T$, $\piNN_{\text{non rob}}$ becomes more similar to $\pi^{\text{ref}}$.
    }
    \label{fig:ref strategy small trading costs long and short T}
\end{figure}
\paragraph{The importance of market characteristics for non-robust optimal investments} 

So far in this section, we found that across all considered market settings, NN-based non-robust strategies $\piNN_{\text{non rob}}$ that do \emph{not} mimic asymptotically optimal investments $\pi^{\text{ref}}$ nonetheless perform on par with those asymptotically optimal reference strategies in terms of $E_{\tilde u_p}$. Moreover, we found that, interestingly, the training of NN-based strategies leads to reduced VaR of the corresponding terminal portfolio value distributions (cp. \Cref{tab:results small tans costs}).  
For non-robust utility optimization in discrete time, real-world scenarios, this implies that there are suitable trading strategies beyond the classically known asymptotic strategy. Those alternative strategies should potentially be preferred if reducing VaR is of interest as well.

Naturally, the question arises about what really drives the differences between NN-based, numerically optimized strategies $\piNN_{\text{non rob}}$ on the one hand, and asymptotically optimal strategies $\pi^{\text{ref}}$ on the other. In a series of experiments, we found that clearly, the magnitude of drift and diffusion of reference markets in combination with the time horizon matters for optimal investments in discrete-time settings. 

NN based strategies are able to adapt to concrete real-world settings and learn to invest more in risky assets, only if drift and diffusion characteristics allow for more potential upside over the dedicated trading time horizon, compared to the risk-less bank account investment. Depending on the time horizon, $\piNN_{\text{non rob}}$ might look very different even if annual drift and diffusion characteristics are held at the same level. The asymptotic strategy $\pi_{\text{ref}}$ however, keeps proportional investments in the no-trade region corresponding to those annual characteristics, irrespective of the time horizon. For reasonable configurations of market characteristics, the difference between the NN-based $\piNN_{\text{non rob}}$ and the asymptotic solution $\pi^{\text{ref}}$ is therefore most prominent in short-term trading.

More concretely, the risky asset $S$ has an average return of $e^{\mu_S \Delta t}$ over a time period $\Delta t$. If the reference strategy $\pi^{\text{ref}}$, which does not depend on the time horizon $T$, suggests to buy a certain amount $H_0$ of stocks at $t=0$, then this leads to trading costs of $H_0 S_0 c_{\text{prop}}$, irrespective of the time the stocks can be held. Therefore, as soon as $T$ is so small that the expected discounted profit is smaller than the trading costs, $H_0 S_0 (e^{\mu_S T}-1) < H_0 S_0 c_{\text{prop}}$, i.e., when $e^{\mu_S T}-1 < c_{\text{prop}}$, the investment into the risky asset has a negative expected profit incorporating trading costs. 
While $\pi^{\text{ref}}$ does not take this into account, the NN strategy $\piNN_{\text{non rob}}$  learns when it is beneficial not to invest into the risky asset at all, due to this trading cost effect. In particular, we see in experiments with small $T$ that $\piNN_{\text{non rob}} \approx 0$ (\Cref{fig:ref strategy small trading costs long and short T} left)\footnote{Here, $e^{\mu_S T}-1 \approx 0.0045 < c_{\text{prop}} = 0.01$.}, while it becomes more similar to  $\pi^{\text{ref}}$ for larger $T$ (\Cref{fig:ref strategy small trading costs long and short T} right)\footnote{
It is important to note that not all paths of $\piNN_{\text{non rob}}$ are as similar to $\pi^{\text{ref}}$ as the plotted one. For some paths, $\piNN_{\text{non rob}}$ lies above and for some it lies below the reference strategy. However, on average they seem to behave similarly.
}\textsuperscript{,}\footnote{In \Cref{fig:ref strategy small trading costs long and short T}, we used step sizes of $1/650$ and $1/130$ for $T=0.1$ and $T=2$, respectively.}. 
For intermediate levels of $T$, as in the experiments in \Cref{fig:ref strategy small trading costs}, $\piNN_{\text{non rob}}$ seems to choose a type of interpolation between the behaviours. 

\begin{figure}[!t]
    \centering
    \includegraphics[width=.49\textwidth]{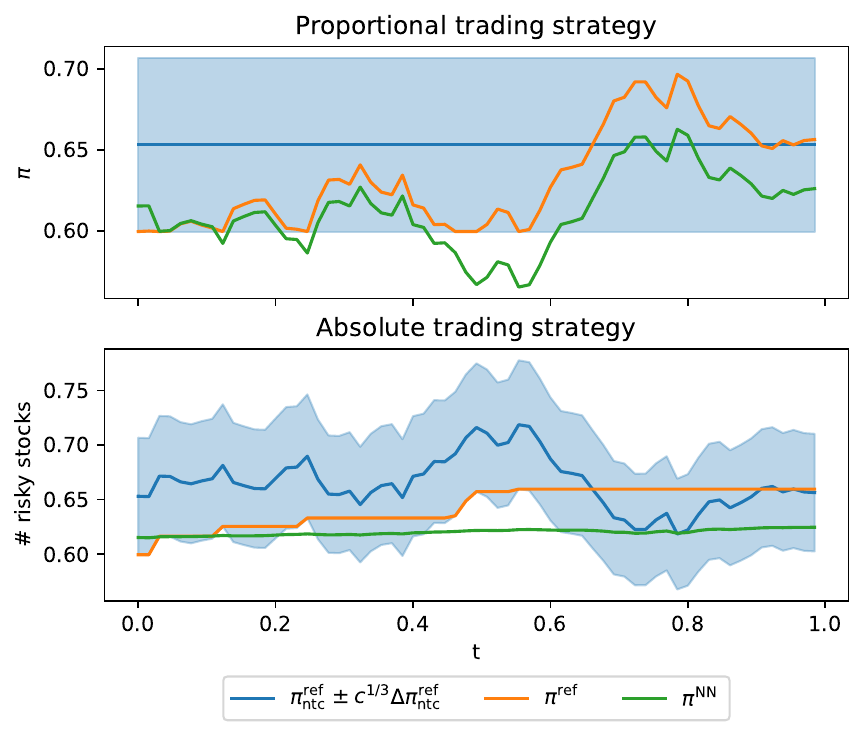}
    \includegraphics[width=.49\textwidth]{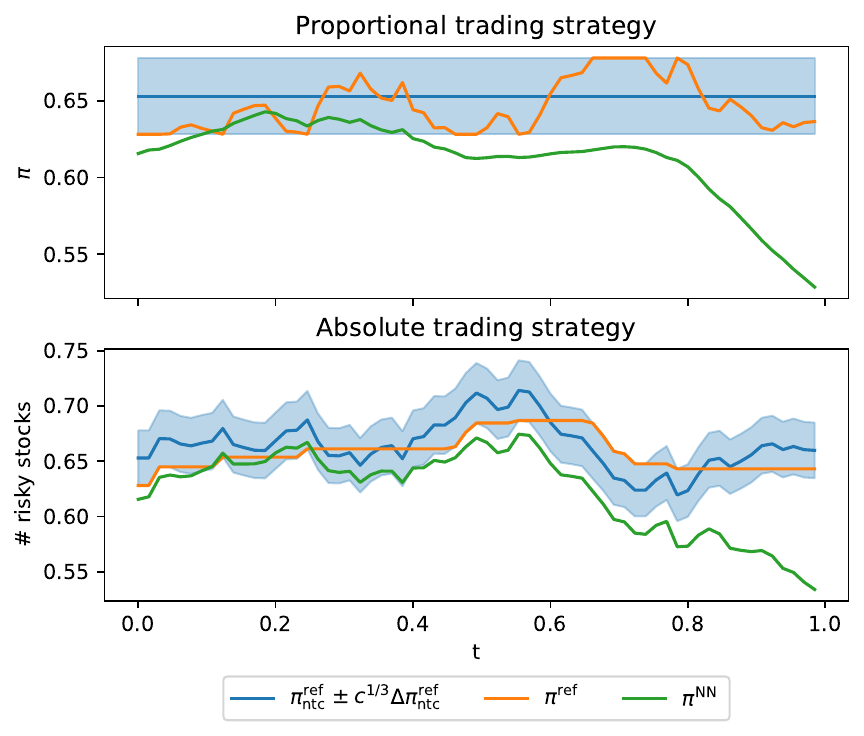}
    \caption{
    Test paths of the optimal reference strategy without trading costs $\pi_{\text{ntc}}^{\text{ref}}$ together with the no-trade-region, the asymptotically optimal reference strategy with trading costs $\pi^{\text{ref}}$ and of the non-robustly trained GAN strategy $\piNN_{\text{non rob}}$ for $T=1$ and $c_{\text{prop}} = 0.1\%$ (left) and $c_{\text{prop}} = 0.01\%$ (right) in the market with $\tilde \mu =5.5\%$ and $r=1.5\%$.
    }
    \label{fig:ref strategy small trading costs smaller costs}
\end{figure}
When decreasing the trading costs $c_{\text{prop}}$, we see a similar behaviour as when increasing $T$. Reducing  $c_{\text{prop}}$ from $1\%$ to $0.1\%$ causes the NN strategy $\piNN_{\text{non rob}}$ to become more similar to $\pi^{\text{ref}}$ (\Cref{fig:ref strategy small trading costs smaller costs} left). This is expected because the expected profit incorporating trading costs increases with decreasing cost, as is the case for growing $T$.
Further decreasing the costs, $\piNN_{\text{non rob}}$ starts to resemble $\pi_{\text{ntc}}^{\text{ref}}$ (particularly visible in terms of $H$), with a tendency to sell stocks towards the end of the time horizon (\Cref{fig:ref strategy small trading costs smaller costs} right).
The opposite behaviour happens for increasing trading costs (being similar to decreasing $T$), where a critical cost level exists from which onwards it is not optimal to transact because the expected profit incorporating trading costs is negative. This was already found by \citet{Atkinson2010Building} (and is also implied by the structure of the no-trade region \eqref{equ:no trade region} of \citet{muhlekarbe2017primer}). However, the authors did not draw the connection to the time horizon $T$, i.e., that the expected profit incorporating trading costs also depends on the maximal time to hold a stock and hence, that the time horizon influences whether or not it is optimal to trade.

\begin{table}[!tb]
\caption{
Numerical results of reference strategy and trained GAN strategies evaluated against the reference market with $\tilde\mu=5.5\%$, $r=1.5\%$ by \eqref{eq:explicitSigExpectedUtility}.
Additionally, the discounted value at risk $\operatorname{VaR}_\alpha$ for $\alpha = 5\%$ is computed in the noise-free reference markets. The standard choice is $T=1$ and $c_{\text{prop}}=1\%$.
}
\label{tab:results small tans costs for small and large T}
\begin{center}
\resizebox{\textwidth}{!}{
\begin{tabular}{l | r r | r r | r r | r r}
\toprule
\multirow{2}{*}{}&\multicolumn{2}{c|}{$T=0.1$}&\multicolumn{2}{c|}{$T=2$}&\multicolumn{2}{c|}{$c_{\text{prop}}=0.1\%$}&\multicolumn{2}{c}{$c_{\text{prop}}=0.01\%$}\\
\cmidrule{2-9}
   & $E_{\tilde u_p}$ & $\operatorname{VaR}_\alpha$ & $E_{\tilde u_p}$ & $\operatorname{VaR}_\alpha$ & $E_{\tilde u_p}$ & $\operatorname{VaR}_\alpha$ & $E_{\tilde u_p}$ & $\operatorname{VaR}_\alpha$ \\ \midrule[\heavyrulewidth]
$\pi^{\text{cash}}$ & {\bf 2.0015} & 0 & 2.0302 & 0 & 2.0151 & 0 & 2.0151 & 0 \\
$\pi^{\text{ref}}$    & 1.9973 & 0.10 & {\bf 2.0506} & 0.37 & {\bf 2.0277} & 0.30 & {\bf 2.0283} & 0.31   \\
$\piNN_{\text{non rob}}$      & {\bf 2.0015} & 0  & {\bf 2.0500} & 0.35 & {\bf 2.0276} & 0.28 & {\bf 2.0282} & 0.29 \\
\bottomrule
\end{tabular}
}
\end{center}
\end{table}
In \Cref{tab:results small tans costs for small and large T}, we see that the learned strategies $\piNN_{\text{non rob}}$ perform on par with $\pi^{\text{ref}}$ in terms of expected utility, while again attaining smaller $\operatorname{VaR}_\alpha$, in all cases for different time horizon $T$ and trading costs $c_{\text{prop}}$. 
In particular, in all cases, the NN learns an alternative strategy (as can also be seen in the plots) that has (at least) comparable performance to the reference strategy, but does not exhibit the behaviour of no-trade regions.

\subsubsection{Comparison in Markets with Heavy Tailed Returns}\label{subsubsec:Comparison in Markets with Heavy tails}
In this section, we leverage the flexibility of our GAN-based algorithm to learn optimal investments in markets with heavier tails. To that end, we consider a market with risk-free asset $B_t = e^{rt}$ and one risky asset $S$, whose log-returns are independent, student-t distributed with $\nu$ degrees of freedom, location $\tilde\mu$ and scale $\tilde\sigma$. In such markets, no explicit solutions for optimal investments exist so far, even in non-robust settings for vanishing transaction costs. We therefore compare those robust and non-robust strategies to the asymptotically optimal reference strategy from \Cref{sec:Comparison of the Reference and the GAN Strategy}.

\paragraph{Experiment setup}
In our experiments of this section, we use a similar setup as in \Cref{sec:Comparison of the Reference and the GAN Strategy} with $p=0.5$, $c_{\text{prop}} = 1\%$, $X_0 =1$, $S_0=1$, $\tilde\sigma=35\%$, $\delta_t = 1/65$ with $65$ steps. We set the drift and risk-free rate $(\tilde\mu_S, r) = (7\%, 3\%)$, hence, $\tilde\mu = 10\%$, and consider two different regimes for the degrees of freedom $\nu\in\{3.5, 20\}$.
The robust training, model validation and model evaluation is analogous to the one explained in the experiment setup of \Cref{sec:Comparison of the Reference and the GAN Strategy}.

\begin{table}[!tb]
\caption{
Numerical results of reference strategy and trained GAN strategies evaluated against the reference market with $\tilde\mu=10\%$, $r=3\%$ and independent student-t returns with $\nu$ degrees of freedom via \eqref{eq:explicitSigExpectedUtility}.
Additionally, the discounted value at risk $\operatorname{VaR}_\alpha$ for $\alpha = 5\%$ is computed in the noise-free reference markets.
}
\label{tab:results small tans costs student t}
\begin{center}
\begin{tabular}{l | r r r | r r}
\toprule
\multirow{2}{*}{}&\multicolumn{3}{c|}{$\nu=20$}&\multicolumn{2}{c}{$\nu=3.5$}\\
\cmidrule{2-6}
   & $\displaystyle M_{\tilde u_p}(\pi)$ & $E_{\tilde u_p}(\pi, \tilde\mu, \tilde\sigma)$ & $\operatorname{VaR}_\alpha$ &  $E_{\tilde u_p}(\pi, \tilde\mu, \tilde\sigma)$  & $\operatorname{VaR}_\alpha$ \\ \midrule[\heavyrulewidth]
   $\pi^{\text{cash}}$
& \multicolumn{2}{c}{$2.0302$} & 0 &{$2.0302$} & 0 \\
$\pi^{\text{ref}}$ & 2.0139 & {$ 2.0576$} &$0.49$& $-\infty$& -\\
$\piNN_{\text{non rob}}$  & 2.0264  & {\bf 2.0583} &$0.41 $& {\bf 2.0430} & $0.26$\\
$\piNN_{\text{fully rob}}$ &  {\bf 2.0385} & $2.0515$& $0.22 $& $-\infty$ & - \\
\bottomrule
\end{tabular}
\end{center}
\end{table}
\begin{figure}[!b]
    \centering
    \includegraphics[width=.49\textwidth]{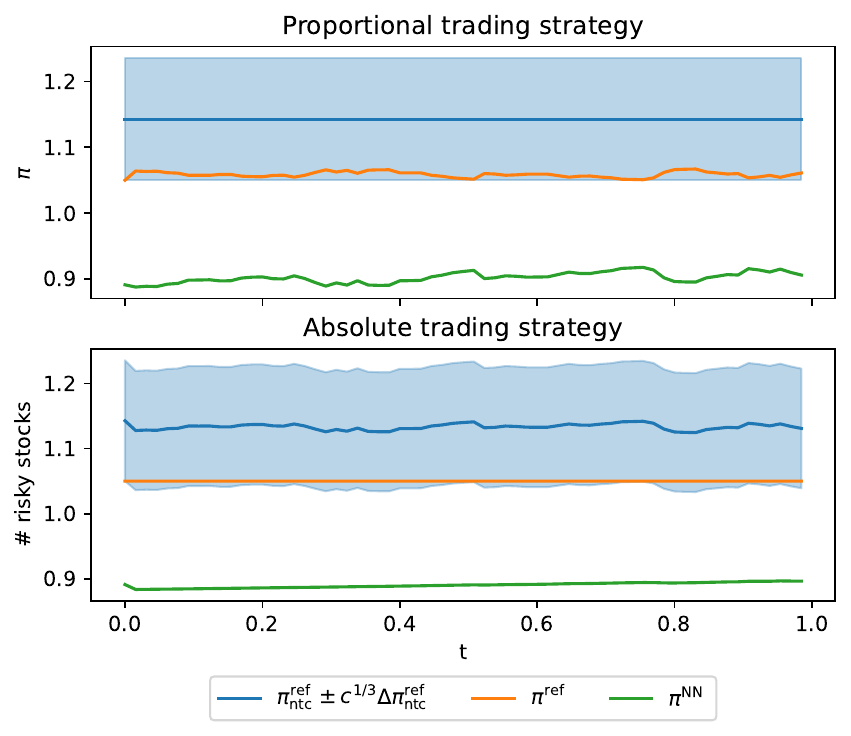}
    \includegraphics[width=.49\textwidth]{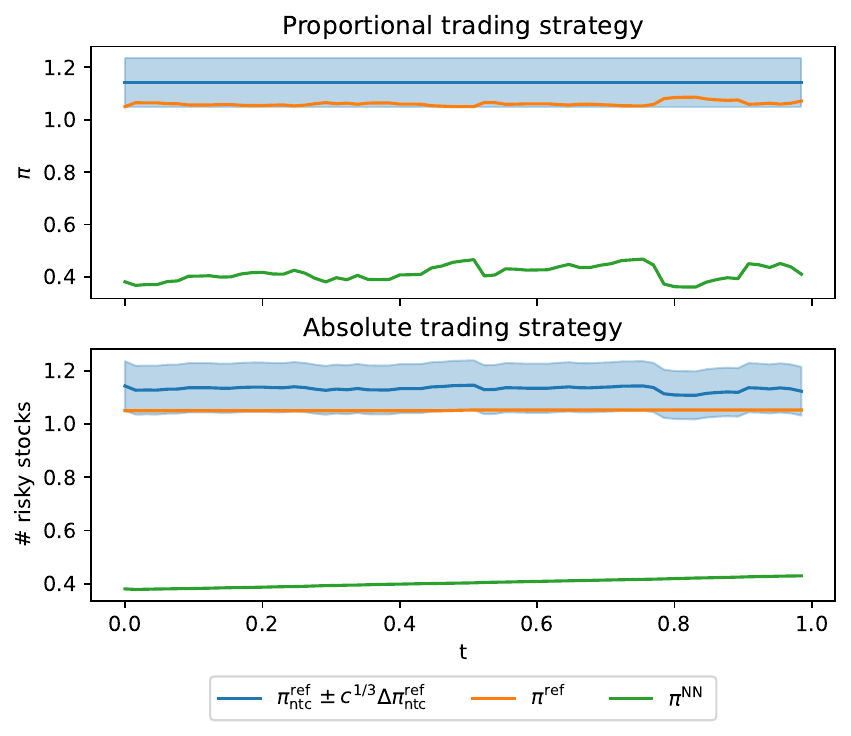}
    \caption{
    Test paths of the optimal reference strategy without trading costs $\pi_{\text{ntc}}^{\text{ref}}$ together with the no-trade-region, the asymptotically optimal reference strategy with trading costs $\pi^{\text{ref}}$ (optimal only w.r.t.\ the Black--Scholes market), and of the non-robustly trained GAN strategy $\piNN_{\text{non rob}}$. The left plot shows paths in the market with $\nu=20$, and the right plot with $\nu=3.5$.
    In each of the two sub-figures, the proportional ($\pi_t$; top) and the absolute ($H_t$; bottom) strategies are plotted.
    For the reference strategy we see that $H_t^{\text{ref}}$ stays constant within the no-trade region, which implies a changing portfolio weight $\pi^{\text{ref}}_t$, due to the change in price $S_t$ and portfolio value $X_t$. 
    }
    \label{fig:ref strategy small trading costs student t}
\end{figure}
\paragraph{Results}
In \Cref{tab:results small tans costs student t}, we show a comparison of the reference strategy $\pi^{\text{ref}}$ and the GAN-based strategies $\piNN$ evaluated against the reference markets (with highest, i.e., best values per column in bold). For the fully robust strategy $\piNN_{\text{fully rob}}$, we show the results of the best combination of scaling factors, given by $\lambda_1=1, \lambda_2=10$ (for $\nu=20$).

The market with $\nu=20$ degrees of freedom closely resembles the corresponding Black--Scholes market from \Cref{sec:Comparison of the Reference and the GAN Strategy}. Accordingly, the GAN-based strategies in that market perform similarly as before: $\piNN_{\text{non rob}}$ performs comparably to $\pi^{\text{ref}}$ in terms of expected utilities  $E_{\tilde u_p}$, with the additional benefit of reducing risk (reduced $\operatorname{VaR}_\alpha$). 
For GAN-based, fully robust strategies $\piNN_{\text{fully rob}}$, this effect intensifies, yielding strategies that  reduce $\operatorname{VaR}_\alpha$ even further, at the expense of lesser expected utility $E_{\tilde u_p}$ on the reference market. 
However, when evaluated against the noisy market with $M_{\tilde u_p}$, $\piNN_{\text{fully rob}}$ outperforms all other strategies.
Moreover, all strategies outperform the pure bank account investment $\pi^{\text{cash}}$ in $E_{\tilde u_p}$.

In the market with $\nu=3.5$ degrees of freedom, we observe that the asymptotic strategy $\pi^{\text{ref}}$ leads to defaults (i.e., negative portfolio value) at some time during trading in the evaluation. The loss incurred at default leads to negative terminal portfolio values, whose utility $\tilde{u}_{p}$ is not well defined. In \Cref{tab:results small tans costs student t}, we indicate negative infinity for values of $E_{\tilde u_p}$ for those strategies that can lead to defaults with non-zero probability (i.e., on at least one sample path during evaluation).  Note that also with the fully robust GAN-based strategy $\piNN_{\text{fully rob}}$, we incur defaults at some point during its training. We suspect, however, that with more careful hyper-parameter selection in the GAN training, the robust strategies would not face this issue.

While  $\pi^{\text{ref}}$ and  $\piNN_{\text{fully rob}}$ (without hyper-parameter optimization for the novel market setting) struggle to avert defaults in all market scenarios, the non-robustly trained $\piNN_{\text{non rob}}$ manages to trade without default on all scenarios even in the heavy tailed market with $\nu=3.5$ degrees of freedom. Moreover, we see in \Cref{tab:results small tans costs student t} that $\piNN_{\text{non rob}}$ outperforms the pure bank account strategy in terms of $E_{\tilde u_p}$, while keeping the VaR at a low level of $26\%$. This level is comparable to the one of the fully robust strategy in the corresponding Gaussian market (cp. $\operatorname{VaR}_\alpha(\piNN_{\text{fully rob}})=0.25$ in \Cref{tab:results small tans costs}).

In \Cref{fig:ref strategy small trading costs student t}, we show paths of the non-robustly trained GAN-based strategy in comparison to the asymptotically optimal solution of \Cref{sec:Comparison of the Reference and the GAN Strategy}, on one trajectory of the heavy tailed risky asset with $\nu=20$ (left) and $\nu=3.5$ (right) degrees of freedom, respectively. While the absolute strategies (bottom plots) develop similarly in both markets, with gradually increasing numbers of risky assets over time, the GAN-based strategy learned to adapt to the market environment by investing about half as much in the risky asset in the market with $\nu=3.5$ as in the market with $\nu=20$.
This highlights one of the main strongholds of our GAN-based algorithm: its robustness towards changing market environments and its flexibility to yield optimized investments even under these changing conditions.

\section{A Note on the Path-Dependence of the Trading and Market Strategy}
In this section, we discuss in more detail the path-dependence of the optimal trading strategy and market characteristics in the context of penalized robust utility optimization with path-dependent penalties and transaction costs. Trading costs and path-dependent penalties are the two factors increasing the complexity of the robust utility optimization problem such that a (Markovian) explicit solution as in \Cref{sec:Setting with explicit Solution} can no longer be expected.

\subsection{Path-Dependent Penalties}
Considering the path-dependent penalties introduced in \Cref{sec:RealisticSettingForStockMarket}, it becomes apparent that the optimal robust diffusion characteristic is path-dependent if $\sigma^\phi_n \ne 0$ for all $n$. To see this, notice that $\hat l_{\text{QCV}}(\ln(S))$ of \Cref{eq:QuadCovarApprox} depends on all time steps of the realized stock path $S$, even though it is an approximation of the penalty $l_{\text{QCV}}(\ln S) = [\ln S]_T = \int_0^T \Sigma_s ds$, which depends only on the volatility $\Sigma$. 
Since the realized stock path $S$ depends on the random increments of its driving Brownian motion, it cannot be reconstructed from the (past) market characteristics and the current value of $S$. Therefore, the diffusion characteristic optimising the penalty $\hat l_{\text{QCV}}$ necessarily needs access to the entire history of the path of $S$. Indeed, any Markovian $\sigma^\phi \ne 0$, has positive probability of exceeding the given reference value  $\hat l_{\text{QCV}}(\ln S^{\tilde\mu, \tilde\Sigma})$ at any time $t_n$ (due to the unbounded Gaussian increments). 
Without the information of the path of $S$, Markovian characteristics are, in general, not able to know whether the current value of the penalty $\sum_{i=1}^n(\ln S_{t_i}- \ln S_{t_{i-1}})^2$ actually already exceeded the reference value\footnote{In some situations a Markovian strategy knowing $S_0$ can know this solely based on the current value of $S$, e.g., if this value is so large that even the smallest possible steps to reach it within the given time steps leads to a penalty value larger than the reference value. But this is not true in general, since already in the next step $S$ could be (close to) $0$ again.}. By contrast, path-dependent diffusion characteristics having access to the path of $S$, also can access the current value of the penalty. Therefore they can decide whether to set $\sigma^\phi_k \approx 0$ for $k > n$, if the reference value is already exceeded, or to a larger value otherwise.
This gives intuition on why the optimal robust market characteristics are path-dependent, at least when the major objective for the market is to satisfy this penalty, i.e., if the corresponding scaling factor $\lambda_1$ is large.
In the limit case $\lambda_1 \to \infty$, this also leads to a path-dependent optimal trading strategy\footnote{In the setting with 1 risky asset we have $\ln S_{t_i}- \ln S_{t_{i-1}} = \mu^\theta_{i-1} \Delta t_i + \sigma^\phi_{i-1} \Delta W_{t_i}$, therefore, the market will set $\mu^\theta = 0$ as soon as the reference value $\hat l_{\text{QCV}}(\ln S^{\tilde\mu, \tilde\Sigma})$ is exceeded. Consequently, the trading strategy should allocate everything in the risk  free asset, as soon as this happens.}. 
We note, however, that our early stopping validation metric and the evaluation metric do not consider the penalties of the learned market strategy and solely depend on the learned trading strategy. Therefore, path-dependent trading strategies are not necessarily advantageous. This goes in line with our findings that there is no improved performance when considering path-dependent strategies. 

\subsection{Transaction Costs}
The asymptotically optimal trading strategy for small transaction costs, which was discussed in \Cref{sec:Robust Utility Optimization with Small Trading Cost}, has a generic path-dependence in terms of the inputs that we use. In particular, it needs knowledge of the amount of stocks $H_{t_{n-1}}$ held at the previous time step $t_{n-1}$ to decide whether to keep the same number of stocks or to move with the boundary of the no-trade region. While the strategy could be made Markovian by including $H$ as an input (state) variable for the strategy, our setting only uses $S,X$ as such. Clearly, $H$ can be reconstructed from the history of the path of $S$ and $X$ such that the path-dependent RNN strategy could in principle learn the asymptotically optimal strategy, while this is impossible for the Markovian FFNN strategy.
Nevertheless, we saw that the asymptotically optimal strategy is not always optimal in settings with small transaction costs, in particular when the time horizon is small. In this case, the optimal strategy is Markovian and simply keeps everything in the risk free asset. In cases where the time horizon is larger, it is not clear what the true optimal strategy looks like.
Our results showed that the learned Markovian strategies achieve the same results in terms of expected utility, while using a qualitatively different strategy.
This suggests the hypothesis that Markovian strategies are dense in the set of all strategies, where the pseudo-distance between strategies is measure in terms of difference in expected utility.

\section{Conclusion}\label{sec:Conclusion}

In this paper, we have tested a GAN-based approach for penalized robust utility optimization in realistic, real-world settings. 

For problems with instantaneous penalties on drift and diffusion characteristics, we substantially extended on prior work by \cite{Guo2022}, by suggesting a slightly altered algorithm and testing it in a series of experiments. We showed that our GAN-based algorithm recovers known explicit solutions in friction-less market settings. Moreover, we found that with trading costs, GAN-based robust strategies outperform explicit strategies that were obtained as solutions in the friction-less setting. Therefore, GAN-based robust strategies are a valuable alternative to closed form solutions obtained in idealized settings.

We further suggested generic, path-dependent penalties for penalized robust utility optimization and tested our GAN-based algorithm for it.
We proposed several metrics for assessing out of sample robustness of trained strategies based on a pool of potential market measures. Based on these metrics, we found that under proper hyper-parameter optimization, fully robust GAN-based strategies outperform non-robust GAN-based strategies, as well as pure bank account investments. 
Our evaluation showed that the proposed GAN algorithm is feasible in generic settings that lack closed-form or otherwise computable solutions.
Moreover, we saw comparable performance with feed-forward and recurrent NN architectures, suggesting that potentially, optimal strategies to robust utility optimization with path-dependent penalties are not path dependent themselves.

Lastly,  we extended on a prior study of small trading costs in (non-robust) utility optimization by \cite{muhlekarbe2017primer}, comparing their results to our GAN-based robust and non-robust trading strategies. We found that, interestingly, even non-robustly trained GAN-based strategies displayed robust behavior, meaning that they learned to overall reduce risky investments compared to asymptotically optimal investment strategies, while achieving similar expected utility.
This finding suggests that in real-world, discrete-time settings, it might be beneficial to invest less into risky assets than suggested by idealized continuous-time, asymptotic solutions and to neglect no-trade regions.

Overall with this paper, we found that while realistic robust utility optimization is a challenging task, GAN-based algorithms offer feasible, generic approaches to approximate otherwise unknown solutions.
Going forward, it will be particularly interesting to further delve into the optimization with realistic, path-dependent penalties and employ our GAN-based algorithm to uncover potential characteristics of unknown, explicit solutions.

\section*{Declaration of funding}
No funding was received.

\section*{Disclosure of interest}
There are no potential conflicts of interest.

\bibliographystyle{plainnat}
\bibliography{refs}

\newpage
\appendix
\section*{Appendix}

\section{Additional Results for Experiments on Instantaneous Penalizations}
In this section, we report on additional experiments on penalized robust utility optimization with penalties as in \Cref{eq:GuoPenalty}. Penalties of this type tie instantaneous drift and diffusion characteristics $\mu_t$ and $\sigma_t$ to constant reference values $\tilde\mu$ and $\tilde\sigma$, respectively. A first GAN-based approach for penalized robust utility optimization of this type has been introduced in \citet{Guo2022}. We extended their first experimental analyses with experiments on multi-dimensional markets, full robustification and incorporation of transaction costs in \Cref{sec:Setting with explicit Solution,sec:MarketWithTransactionCosts}. 
In the following, we report on further results.

\subsection{Additional Evaluation Metrics and Benchmark Strategies}
As in \Cref{sec:Setting with explicit Solution}, we consider the relative error $\err_{rel}^{\Sigma^\star, \mu^*}(\piNN, \pi^\star)$ that indicates how well the penalized robust utility optimization problem is solved by a learned strategy $\piNN$.
However, the relative error does not tell how well the investment strategy would work in practice, where the market usually experiences some randomness around the reference value $\tilde \Sigma = \tilde \sigma \tilde \sigma^\top$. 

In \Cref{sec:Evaluation Criterion: Early Stopping and General Assessment of Robustness}, we introduced a number of approaches for modeling noisy market models for purposes of model evaluation and model selection. In this section, we revisit the ``constant noise'' approach of \Cref{sec:Evaluation Criterion: Early Stopping and General Assessment of Robustness}.
This approach extends the idea of \cite{Guo2022} to consider a noisy (to the trader unknown) volatility matrix $\sigma^{noisy} = \tilde \sigma + \epsilon \tilde{\sigma}_{noise}$, where the noise has i.i.d.\ entries $(\tilde{\sigma}_{noise})_{i,j} \sim N(0,1)$ and $\epsilon > 0$ is the noise scaling parameter. The corresponding covariance matrix is $\Sigma^{noisy} := \sigma^{noisy} (\sigma^{noisy})^\top$. Similarly, we introduce a noisy drift vector $\mu^{noisy} = \tilde \mu + \epsilon \tilde{\mu}_{noise}$, where the noise has i.i.d.\ entries $(\tilde{\mu}_{noise})_{i,j} \sim N(0,1)$. We consider $\mu^{noisy}$  whenever the drift is also robustified (otherwise we use the fixed drift $\mu^{noisy} = \mu$). 

As in \Cref{sec:Experiments}, we compute expected utilities, now based on these noisy market parameters
\begin{equation*}\label{eq:noisySigExpectedUtility}
    \Ereal(\pi, \Sigma^{noisy}, \mu^{noisy})= \E\left[ u(X_T^{\pi, \mu^{noisy}, \Sigma^{noisy}}) \right],
\end{equation*}
for the learned strategy $\pi=\piNN$ and for $\pi=\pi^*$.
Note that in comparison to \citet{Guo2022}, we do not only use one random sample for the noisy market parameters, but generate $1000$ i.i.d.\ samples. 
Then, we compute the minimum over all samples (which approximates  $\inf_{\P\in\cP} \E_{\P}\left[u(X_T^\pi)\right]$) as well as the mean, median, standard deviation and the $95\%$ approximate normal confidence interval. 
We compare those metrics for $\pi = \piNN$, with the ones for the explicit solution $\pi = \pi^*$.

Moreover, we compare these values to the metrics for the optimal investment strategy $\piOracle_i :=(\Sigma^{noisy, (i)})^{-1} (\mu^{noisy} -r)$. 
This reference strategy corresponds to an ``oracle'' solution, where the full (but usually not available) information about the market is used to get the maximal possible expected utility (by solving \eqref{eq:explicit system 1} or \eqref{eq:explicit system 2}  with the knowledge of $\Sigma^{noisy}_i$ and $\mu^{noisy}$ for $\pi$).
Clearly, this should only be considered as an upper bound, but not as a reachable goal, since exact information about the market is used, which is usually not available. \\

Moreover, as an additional reference value, we compute those metrics with the investment strategy $\piNN_{nr}$ trained to optimize the non-robust optimization task
\begin{equation}\label{eq:NonRobustMultiDObjective}
    \sup_\pi \E\left[ u(X_T^{\pi, \tilde\mu, \tilde{\Sigma}}) \right],
\end{equation}
i.e., where only the generator is trained (without a discriminator) and $\tilde \Sigma$ is used as fixed volatility matrix together with the fixed drift $\tilde{\mu}$. 
This corresponds to penalty scaling factors $\lambda_{i} = \infty$.

\subsection{Additive and Multiplicative Penalization for the Volatility}\label{app:Multiplicative Penalization for the Volatility}
 We define the \emph{additive} penalty function appearing in  \Cref{eq:GuoPenalty,eq:additive instant penalty}
\begin{equation*}
    F_a(\Sigma_t):=\lVert \Sigma_t  - \tilde  \Sigma \rVert_F^2,
\end{equation*}
where $\Sigma_t = \sigma_t \sigma_t^\top$.
Moreover, we additionally consider the multiplicative penalization 
\begin{equation}\label{eq:multiplicative instant penalty}
    F_m(\Sigma_t) = \lVert \Sigma_t  \tilde \Sigma^{-1} - I \rVert_F^2.
\end{equation}
 An explicit solution of the robust utility optimization problem with instantaneous penalties building on $F_m$ can be derived along the same lines as in \Cref{sec:Setting with explicit Solution}, by solving the equation system
\begin{equation}\label{eq:explicit system 2}
    \Sigma = \frac{\pi \pi^\top \tilde\Sigma \tilde\Sigma }{4 \lambda_1} + \tilde \Sigma \quad \text{and} \quad \Sigma \pi = (\mu - r).
\end{equation}

\subsection{Settings with Explicit Solution}\label{app:Setting with explicit Solution}

\subsubsection{Additional Results: Stock Market with 2 Uncorrelated Risky Assets -- Robust Volatility Case}\label{sec:Additional Results: Stock Market with 2 Uncorrelated Risky Assets -- Robust Volatility Case}
In this section, we analyze additional metrics for the settings (S) and (AS) of \Cref{sec:StockMarketwithMultipleRiskyAssets}.

\paragraph{How well does the GAN algorithm solve the optimization problem?}
We observed in \Cref{sec:StockMarketwithMultipleRiskyAssets}, that for additive penalties involving $F_a$, the learned strategies $\piNN$ achieved $\err_{rel}^{\Sigma^\star, \mu}(\piNN, \pi^\star) < 0.2\%$ in both settings (S) and (AS). Moreover, the GAN training solves the optimization problem \eqref{eq:MultiDObjective} similarly well also for the multiplicative penalization involving $F_m$ with $\err_{rel}^{\Sigma^\star, \mu}(\piNN, \pi^\star) < 0.2\%$ in all tested hyper-parameter combinations.

\paragraph{How does the robust strategy perform in a noisy market?}
We test two noise levels $\epsilon \in \{ 0.025, 0.15 \}$.
For both (S) and (AS), both noise levels and both penalization functions $F_a$ and $F_m$, average expected utilities based on noisy market volatilities $\sigma^{noisy}$ for learned strategies and for explicit strategies are indistinguishable (on a 5\% level of confidence).

For $\epsilon=0.025$, i.e., for low noise, we observe that the non-robustly trained strategies $\piNN_{nr}$ perform best in terms of average expected utilities based on noisy market volaitilies. This is in line with the results in \citet[Figures 1-3]{Guo2022}, where non-robust strategies show higher utilities than robust ones for small noise levels. When increasing the noise, the robust strategies yield higher average utilities than non-robust ones. Intuitively, this behaviour is very natural; the larger the noise, the more the noisy volatilites $\sigma^{noisy}$ differ from the reference volatility $\tilde\sigma$, and thus, the more the robust optimization pays off.

\paragraph{How close are the trained strategies to the gold standard?}
The oracle solution leads to an average expected utility that is significantly (on a 5\% confidence level) higher than the one of either the learned or the explicit strategies.
In particular, this additional information about the market leads to a significant advantage that manifests in larger expected utilities. This observation reinforces our choice, not to use the discriminator output as input for  the generator (and vice versa) as it was done in \citet{Guo2022}, since this should lead to unrealistically high expected utility results that can not be reproduced in a real world setting.

\subsubsection{Additional Results: Stock Market with 2 Correlated Risky Assets -- Robust Volatility Case}\label{sec:Additional Results: Stock Market with 2 Correlated Risky Assets -- Robust Volatility Case}
We proceed to analyze additional metrics for the settings (PS), (PAS), (NAS) of \Cref{sec:StockMarketwithMultipleRiskyAssets}.

\paragraph{How well does the GAN algorithm solve the optimization problem?}
We recall that we had small relative errors with the additive regularization functional $F_a$ in \Cref{sec:StockMarketwithMultipleRiskyAssets}.
With the multiplicative regularization functional $F_m$, training the GAN objective \eqref{eq:MultiDObjective} seems harder (or further hyper parameter tuning might be needed): while in (PS), $E^*(\piNN)$ is lower than $E^*(\pi^*)$ on the third decimal place, for (PAS) and (NAS), their difference even ranges within first and second decimal places.

\paragraph{How does the robust strategy perform in a noisy market?}
A similar distinction in performance crystallizes for the noisy market metric, where we use the same noise levels as in \Cref{sec:Additional Results: Stock Market with 2 Uncorrelated Risky Assets -- Robust Volatility Case}. 
When considering the additive regularization $F_a$, across all settings and for both noise levels, $\E_{\Sigma^{noisy}}\ \left[ \Ereal(\piNN, \Sigma^{noisy}) \right]$ for learned strategies is indistinguishable (on a 5\% level of confidence) from $\E_{\Sigma^{noisy}}\left[\Ereal(\pi^*, \Sigma^{noisy})\right]$.
Strategies trained with $F_m$ perform significantly worse in $\Ereal$ than the explicit strategies based on noisy volatilities $\Sigma^{noisy}$. For (PS) and (PAS), this effect seems to weaken with higher noise. In (NAS), even in high-noise scenarios, the robustly learned strategy is outperformed by the explicit strategies in the metric $\Ereal$.

As in the uncorrelated setting, we also observe that in low-noise settings,  $\Ereal(\pi, \Sigma^{noisy})$ is highest for non-robustly trained strategies $\pi$, while in settings with large noise, non-robustly trained $\pi$s result in smallest expected utility $\Ereal(\pi, \Sigma^{noisy})$.

\paragraph{How close are the trained strategies to the gold standard?}
Also for correlated assets, $\E_{\Sigma^{noisy}} \left[ \Ereal(\piOracle, \Sigma^{noisy}) \right]$ is significantly (on a 5\% confidence level) higher than for either the learned or the explicit strategies.

\subsubsection{Additional Results: Stock Market with 5 Uncorrelated Risky Assets -- Robust Volatility Case}\label{sec:Additional Results: Stock Market with 5 Uncorrelated Risky Assets -- Robust Volatility Case}
We analyze the setting with 5 uncorrelated risky assets of \Cref{sec:StockMarketwithMultipleRiskyAssets}.

\paragraph{How well does the GAN algorithm solve the optimization problem?}
As in the uncorrelated 2D setting, the GAN training solves \eqref{eq:MultiDObjective} sufficiently well also for $F_m$ with $\err_{rel}^{\Sigma^\star, \mu}(\piNN, \pi^\star) < 0.2\%$.

Interestingly, we observe that the non-robust training fails to solve the non-robust problem \eqref{eq:NonRobustMultiDObjective}. Even though additional hyper-parameter optimization should resolve this problem, it is interesting that it does not work with the same hyper-parameters that worked well for the robust training. 

\paragraph{How does the robust strategy perform in a noisy market?}
When considering the additive regularization $F_a$, for both noise levels, we have that $\E_{\Sigma^{noisy}}\ \left[ \Ereal(\pi, \Sigma^{noisy}) \right]$ for learned strategies $\piNN$ is indistinguishable (on a 5\% level of confidence) from the explicit strategy $\pi^*$. Moreover, the results of comparing different penalization factors $\lambda_1$ are in line with intuition: for lower noise, $\lambda_1=10$ is the best choice in terms of $\Ereal$, for higher noise, we observe that $\lambda_1=0.1$ leads to highest $\Ereal$.
The results are similar for regularization $F_m$, although it is not so clearly visible which value for $\lambda_1$ is optimal in low noise settings.

\paragraph{How close are the trained strategies to the gold standard?}
Also here, the oracle strategy is significantly (on a 5\% confidence level) better than either the learned or the explicit strategies. (For low-noise scenarios the difference is smaller than for high-noise scenarios.)

\subsection{Additional Results: Stock Market with Proportional Transaction Costs -- Robust Volatility Case}\label{sec:Additional Results: Stock Market with Proportional Transaction Costs -- Robust Volatility Case}
We evaluate the settings of \Cref{sec:StockMarketwithMultipleRiskyAssets} with transaction costs in the noisy market metric.
During the GAN training, we use \Ereal\ on a validation set as lookback criterion for model selection for $\piNN$. This means that we sample a (reasonably large) number of validation paths of the risky assets based on realizations of noisy volatilities and choose $\piNN$ from our training history that maximized a sample estimate of \Ereal\ on these validation paths. As in our test criterion, we consider for these noisy validation paths both the noise levels $\epsilon\in\{0.025, 0.15\}$. This means that at testing time, we are able to compare for each scenario two NN-based strategies: one in which we stop the training based on a noisy estimate of \Ereal\ with the correct market noise, and one where market noise is over- or underestimated.

\subsubsection{Stock Market with 2 Risky Assets}
We consider all settings (S), (AS), (PS), (PAS), and (NAS), in noisy markets with $\epsilon\in\{0.025, 0.15\}$.

Across all these settings, for both high and low transaction costs, and for both regularizing functionals $F_a$ and $F_m$ $\piNN$ outperforms $\pi^*$ in the metric \Ereal\ on a $5\%$ confidence level. 

For high transaction costs in low-noise settings, $\piNN$ still performs as good or better than the constant cash-only strategy, while $\pi^*$ leads to severely lower, sometimes even negative expected utility $\Ereal$.

For both high and low transaction costs, we observe that the non-robustly trained strategies yield higher \Ereal\ than the robustly trained ones in low-noise settings. This is in line with the results from experiments without transaction costs. 

In settings of high noise $\epsilon=0.15$, non-robustly trained strategies perform worse than robustly trained ones, and are also outperformed by the cash-only strategy. Moreover, when in addition transaction costs are high, it becomes increasingly difficult to beat the cash-only strategy.

Finally, as expected, we observe that the NN-strategies selected based on the correct market noise perform generally better than the ones selected based on under- or overestimated market noise. However, only in a  minority of experiments we observe that $\piNN$ selected based on a wrong estimate of market noise performs worse than the explicit strategy.

\subsubsection{Stock Market with 5 Uncorrelated Risky Assets}\label{sec:app:Stock Market with 5 Uncorrelated Risky Assets}
In the market with 5 uncorrelated, symmetric, risky assets, we consider our GAN algorithm with different regularization parameters $\lambda_1\in\{0.1,1,10\}$.

For low market noise, we observe that all models significantly outperform $\pi^*$ in terms of $\Ereal$. However, when trading costs are high, the learned strategies do not perform better than a pure-cash strategy. As in the setting without trading costs, we see that $\lambda_1=10$ leads to highest \Ereal.

Also for high market noise, we observe that GAN models significantly outperform $\pi^*$ in terms of \Ereal\ (except for two models with small $\lambda_1$ where validation was based on small noise).

For small trading costs, the model trained with the smallest choice of $\lambda_1=0.1$ outperforms the cash-only strategy (no matter which noise level was chosen for model selection) when using regularization functional $F_a$. When trading costs are high, both these strategies perform on par with the non-robustly trained strategy $\piNN_{nr}$.

Similarly, when trading costs are high, the model trained with $\lambda_1=0.1$ and regularizing functional $F_m$ performs best (also independent of the validation noise level), however its performance is comparable to the cash-only strategy. For low trading costs, models trained on $F_m$ perform significantly worse than the pure-cash strategy.

Overall, this indicates that the choice of lambda is crucial and needs to be adapted to the markets setting.

\subsection{Additional Results: Robust Drift and other Utility Functions -- No Transaction Costs}
\label{sec:Additional Results: Robust Drift and other Utility Functions -- No Transaction Costs}
We test settings with other utility functions and additionally robustify the drift in settings without transaction costs.

In general, whether or not robustifying the drift is beneficial in terms of \Ereal\ depends on the choice of both $\lambda_1,\lambda_2$, as we have already seen before in \Cref{sec:app:Stock Market with 5 Uncorrelated Risky Assets}. 

\begin{figure}
    \centering
    \includegraphics[width=.7\textwidth]{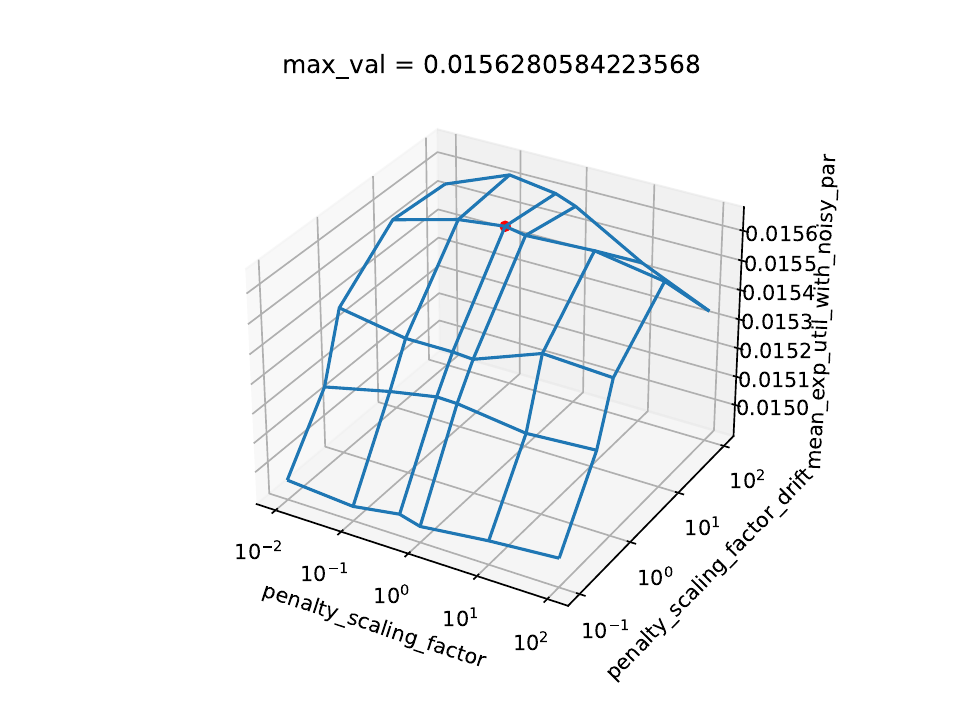}
    \caption{\Ereal\ for NN-strategies trained with 
    robustifying w.r.t.\ $\Sigma$ and $\mu$ for different scaling parameters $\lambda_1,\lambda_2$.}
    \label{fig:Ereal surf plot}
\end{figure}

For the market setting with (AS) and (AD) of \Cref{sec:Robustifying Volatility and Drift} and for the case of power utility with $p=2$, we performed a grid search over $\lambda_1,\lambda_2\in\{0.01, 0.1, 0.5, 1, 10, 100\}$, with results shown in Figure~\ref{fig:Ereal surf plot}, where we evaluated the average expected utility with noisy market drift and volatility. 
The best combination of scaling parameters yields $\E_{\Sigma^{noisy}, \mu^{noisy}} \left[ \Ereal(\pi, \Sigma^{noisy}, \mu^{noisy}) \right] \approx 0.15628$
For comparison, in the non-robust case and in the case where only $\Sigma$ was robustified (corresponding to $\lambda_2=\infty$), the (maximal) values of average  \Ereal\ were $0.01523$ and $0.01556$, respectively.
In particular, robustifying both the volatility and drift is beneficial when choosing the appropriate penalty scaling factors for the given market situation.
Moreover, we see in \Cref{fig:Ereal surf plot} that the calibration of the scaling parameter for the drift is more crucial than for the volatility.

\end{document}